\documentclass[12pt,preprint]{aastex}

\slugcomment{Accepted for publication in ApJ, October 2002}

\shorttitle{Extrasolar planet detection}
\shortauthors{Sparks, W.B.}

\begin{document}


\title{Imaging Spectroscopy for Extrasolar Planet Detection}


\author{William B. Sparks}
\affil{Space Telescope Science Institute, 3700 San Martin Drive, Baltimore, MD
 21218}

\and

\author{Holland C. Ford}
\affil{Dept. of Physics and Astronomy, The Johns Hopkins University, Baltimore
, MD 21218}



\begin{abstract}

We propose
that coronagraphic imaging in combination with moderate to high spectral resolution
from the outset
may prove more effective
in both {\it detecting} extrasolar planets and characterizing them than
a standard coronagraphic imaging approach.
We envisage an integral-field spectrograph coupled to a coronagraph to
produce a datacube of two space dimensions and one wavelength.
For the idealised case where the spectrum of the star is well-known and
unchanging across
the field, we discuss the utility of cross-correlation to seek the extrasolar planet signal,
and describe a mathematical approach to
completely eliminate stray light from the host star (although not
its Poisson noise).
For the case where the PSF is dominated by diffraction and scattering
effects, and comprises a multitude
of speckles within an Airy pattern
typical of a space-based observation,
we turn the wavelength dependence of the PSF
to advantage and present a general way to eliminate the contribution from the
star while preserving both the flux and spectrum of the extrasolar planet.
We call this method ``spectral deconvolution''.
We illustrate the dramatic gains by showing an idealized simulation
that results in a $20$-$\sigma$ detection of a Jovian planet at 2~pc with
a 2-m coronagraphic space telescope, even though the planet's peak flux
is only 1\%\ that of the PSF wings of the host star.
This scales to detection of a terrestrial extrasolar planet at 2~pc with an 8-m coronagraphic
Terrestrial Planet Finder (TPF)
in $\sim 7$~hr (or less with appropriate spatial filtering).
Data on the
spectral characteristics of the extrasolar planet and hence on its atmospheric
constituents and possible biomarkers
are obtained naturally as part of this process.

\end{abstract}




\section{Introduction}

The detection of a large number of extrasolar planets and
planetary systems
through precision radial velocity surveys has revolutionized
thinking on the frequency and characteristics of
planets (Marcy and Butler 1998).
It is now clear that extrasolar planets are common.
Along with the discovery that life can exist and thrive in
an extreme range of environments on Earth, this has given a strong
stimulus and boost to
the search for life elsewhere in the Universe.
Major design studies have been undertaken to consider competing
concepts for the Terrestrial Planet Finder (TPF) whose
primary goal is to locate terrestrial extrasolar planets
in the habitable zone (where liquid water may be present)
and to characterize their atmospheres.
It is thought
that a combination of disequilibrium chemical processes will betray the presence of biological activity
(Woolf and Angel 1998, see also {\it http://tpf.jpl.nasa.gov}).

Because of the faintness of planets, it has generally been assumed that
characterization of Earth-like planets
will be carried out at low spectral resolution.
This
maximizes the number of photons from the planet, per spectral resolution
element, and allows
the use of broad, deep molecular absorption bands
to probe the atmosphere and to seek
evidence of life
(Kasting 1996, Woolf and Angel 1998,
Schindler, Trent \&\ Kasting 2000,
Des Marais et al 2001).
Here, we consider use of higher spectral
resolution observations from the outset, although retaining the imaging context.
The methods described will still work, however, to varying degrees at low
resolution.
Allowing ourselves to consider high resolution enables us to explore
more novel approaches and understand some of the basic dependencies.

High spectral resolution observations have of course been obtained for
the ground-breaking precision radial velocity measurements, and in
addition high spectral resolution
searches have been performed in
order to attempt detection of extrasolar planets through the reflection of
stellar photospheric
emission, coupled to sophisticated numerical methods
(Collier Cameron et al 2002).
Detection of planetary atmospheres with high signal-to-noise
transit spectroscopy has been both proposed and
observed (Brown 2001, Charbonneau et al. 2002).

Classically, direct imaging of faint companions to bright stars has
been carried out with coronagraphic observation.
There has been a great deal of work devoted to improving
the performance of coronagraphs with a multitude of analogue
methods and data analysis techniques
(e.g. Roddier \&\ Roddier 1997, Rouan et al 2000, Spergel and Kasdin 2001;
Kuchner \&\ Traub 2002
and many others).
Other approaches include manipulation of speckle patterns
to reduce PSF haloes or recognize planetary signatures
within the speckle noise, including the dark-speckle
coronagraph (Boccaletti et al 1998) and the use of dichroics
and differencing of speckle images (Racine et al. 1999, Marois et al 2000).

Our goal here is to detect and image extrasolar planets directly,
that is separated from their host star,
through photons reflected or
emitted from their surface or atmosphere.
The concept is to use a coronagraph together with
integral-field spectroscopy (e.g. Bacon et al 1995).
Recognition of the planetary spectrum within the multiple
spectra into which
the image is divided may be achieved using pattern recognition
techniques or by manipulation of the stray light of the host star
with the goal of eliminating it.
The process can be
aided by the orbital velocity of the
planet which displaces its spectrum in wavelength from that of the star.
The concepts outlined may be applicable to either space or ground-based
observation.

Firstly we consider the case where the spectrum of the star is well-known
and consistent across the field of view (i.e. spatially invariant;
the same at all points of interest in its PSF).
In \S~3.1 we
explore the statistical properties of straightforward
cross-correlation of the spectra using a (matched) template.
We show the trade-space between signal-to-noise ($S/N$) and
rejection of stray light from the host star.
In \S~3.2 we present a mathematical approach
which ensures {\it complete} elimination of stray light from the host star.
By using Gram-Schmidt orthonormalization we show that the $S/N$ of the
correlation functions may be preserved while offering formally complete
rejection of the host star light.
In the absence of a matched template, analysis of variance
methods, \S~3.3, may be used to provide information on whether there are
multiple spectroscopic components (i.e. planets or other companions)
present.
We develop spectral templates and idealized simulations for these concepts,
which are more applicable to ground-based observation,
in \S~4.

In \S~5.0 we discuss the alternate case that the PSF of the coronagraphic observation
is dominated by diffraction, or diffraction plus
scattering, and introduce the concept of
``spectral deconvolution''.
This situation is likely to apply to space-based observation, primarily,
although there may be some ground-based opportunities using adaptive
optics.
We use the wavelength dependence of the PSF
to advantage. By changing the image scale inversely to the wavelength 
the PSF is transformed into a simple, slowly varying function that can be
removed e.g. by fitting low order polynomials as a function of
wavelength. The extrasolar planet on the other
hand is moved radially in the datacube by this process, and hence
becomes a high frequency component which is ignored by the starlight
rejection procedure. Reconstruction of the subtracted datacube
back onto the original spatial scale
and summation along the spectral dimension
reveals the extrasolar planet signal.
This method may prove exceptionally powerful and allows essentially
{\it an ideal Poisson-noise limited imaging detection, which simultaneously
provides a spectrum of the extrasolar planet.}
In \S~6, we speculate on future directions a spectroscopic approach
might take, and
\S~7 summarizes our conclusions.

\section{Concept}

Collier Cameron et al (1999), Collier Cameron et al (2002) have utilized sophisticated pattern
recognition techniques to place stringent limits on the contribution 
from a planet to the integrated
light of a star hosting an extrasolar planet.
Tonry \&\ Davis (1979) describe the Fourier theory of cross-correlating
a spectrum with a template and the associated error analysis for finding
the (redshift) peak and (velocity dispersion) width.
Here, we use
elementary concepts involving spectroscopy not only
to characterize extrasolar planets,
but also to assess their utility in the detection process itself.

A standard instrumental approach is to obtain an image of the field of view
which presumably contains a bright star and very faint nearby planet,
perhaps a factor $10^9$ or $10^{10}$ times fainter. Suppression of the starlight
may be achieved through a variety of techniques, most commonly using an
apodized coronagraph.
Observation through several filters yields information on the flux distribution
of the planet with wavelength, and hence its characteristics.

Now suppose we are presented with a similar field of view, but rather
than integrate each portion of the image to measure the total flux from
that region, or pixel, instead we disperse the light in each pixel
using an integral-field spectrograph.
We then explore the resulting datacube of two space dimensions and one spectral dimension
to seek evidence of the extrasolar planet.

Temporal changes changes in spatial PSF structure may be accommodated
since we envisage both spatial and spectral multiplexing.
For long ground-based integrations, we expect rapid fluctations
from the constantly changing speckle pattern to average out in
the spectral dimension. To approach the idealized case of a
spatially invariant spectrum, further low order filtering of
the spectra may be required.
Experimentation with actual coronagraphs and actual
integral field spectrographs would be invaluable in making a determination
of the degree to which such an idealized situation may be approached
in practice.
For space based observation where PSFs are much more stable,
the techniques of \S~5 are more general and likely to
be more easily applied than the pattern recognition methods described
in the earlier sections, although over a limited wavelength
range or close to the host star they may be useful.
While we develop the ideas using both spatial and spectral multiplexing,
either of these may be relaxed at the cost of implicitly imposing
higher stability  demands on the optical system, and additional
exposure time to build up the datacube.

We assume that the dominant noise is the photon noise of the host star,
i.e. that the detector noise is negligible.
We are assuming that the planet is faint and therefore that the photon
budget is dominated by residual light from the host star. If there are $S_0$ photons from
the star in a particular pixel, there are $\approx S_0/(N n_{exp})$
photons
for each of $N$ spectral elements and $n_{exp}$ exposures,
hence this assumption requires
$S_0 > N n_{exp} \sigma_R^2$ if the detector readnoise is $\sigma_R$.
For example, in the simulation of \S~5 below, $S_0 \sim 9\times 10^5$,
$N\sim 600$ so the requirement is $\sigma_R < 38/\sqrt{n_{exp}}$.
For a readnoise of, say, 4~photons, we could take 100~exposures
and satisfy the constraint.

There is an analogy to the principal of ``do no harm''.
By ignoring the spectral
dispersion of the light from each pixel, in our idealized configuration
we simply recover the undispersed coronagraphic observation.
That is, if we add all the counts in each spectrum, we obtain the flux at that pixel.
This represents the best imaging
detection limits currently (and in the near future) achievable.
Yet, with the additional spectral dimension
we may be able to achieve greatly
improved rejection of contaminating starlight,
knowledge of the extrasolar planet's velocity and kinematics,
and knowledge of its atmospheric characteristics,
all with the {\it initial discovery} observation.

The first part of the discussion (\S~3 and \S~4)
considers the use of
pattern recognition techniques to extract information on
the faint extrasolar planet signal within the noisy spectrum under
the assumption that the spectra of the resiudal light from the host star
are the same everywhere in the field of view.
There is a vast array of signal processing and pattern recognition techniques
that could be brought to bear on such a dataset.
To gain insight into some of the fundamental dependencies, however,
we explore simple cross-correlation
of the spectra with a matched template that has
the same shape and structure as the target sought.
The amplitude of the cross-correlation functions, as we show, depends on
essentially two parameters: (i) the total flux from the extrasolar planet relative
to the noise and (ii) the degree of structure,
or variance, of the template.
We extend the analysis to a general analysis of variance, and
conclude this part of the discussion with a technique related to cross-correlation whereby template
spectra are constructed that closely match the extrasolar planet's spectrum but which
are mathematically orthogonal to the stellar spectrum. Cross-correlation
with such a template in the idealized case
fully eliminates stray stellar light, although not
the photon noise from that stray light.

The second part of the discussion (\S~5) offers a completely different
way to remove the stray stellar light for the case where we have strong
wavelength dependent diffraction limited PSFs, e.g. such as
those obtained with space-based observation.
The results of this procedure are {\it both} a Poisson limited image of the
extrasolar planet and, from the outset, its spectrum, irrespective
of the nature of that spectrum.

\section{Correlation and Variance Methods: Theory and Error Analysis}

\subsection{Cross-correlation with matched template}

If the planet has a grey, uniform albedo and isotropic scattering phase function,
then the planet's spectrum is the same as that of the stellar photosphere,
many orders of magnitude weaker, and shifted in velocity by the planet's orbital
motion.
If we are searching for a specific type of extrasolar planet,
such as a terrestrial planet with
strong molecular absorption features in the near infrared,
or a Jovian planet with deep methane troughs,
we may optimize the search
by providing a template whose shape matches what we seek.
We present a variety of templates in \S~4, below.
Eventually, to pursue this approach, one would build a library of spectra for
comparison based on a mix of theory, modelling and observation.

Now consider a pixel in a coronagraphic image of a star and suppose there is a
contribution to the signal from an extrasolar planet at that location.
Let the total flux from the star in the (single) pixel
and within the spectral window be $S_0$ and 
that of the planet be $s_p$.
That is, $S_0$ is the portion of host star flux falling into a single
pixel at some location in a coronagraphic image (not the total
flux of the star over the whole image).
Define a stellar template (spectrum) $T$ such that $\int T(\lambda) dlog\lambda = 1.0$
and $T>0.0$ everywhere (i.e. there are no regions of negative flux).
Since we will sample the spectrum with quantized detector pixels, we
will use a summation version of this, hence $\sum_{i} T_i = 1.0$ where $T_i$
is $T(\lambda_i)$ the template at the $i$-th wavelength $\lambda_i$.
That is, this template is normalized as a probability density
function, and sampled on a logarithmically uniform grid.

The planet will have a spectrum similar in part to that of the star,
since it shines by reflected
starlight, modified by the albedo of the planet as a function of wavelength,
by any absorption features in the planet's atmosphere (through which the reflected
light passes twice), by any intrinsic emission features, such as aurora,
dayglow and nightglow, and shifted by the velocity of the planet.
Let the probability density template $t$ of the planet as a fuction of
(logarithm) wavelength be $t_i$, also such that $\sum_{i} t_i = 1.0$. 

We will use other versions of the templates. $\hat{T}$ is a zero-mean
version of $T$; i.e. $\hat{T} = T - 1/N$ where $N$ is the number of points in the
spectrum.
Also we define $T' = T \times N$;
$T'$ is such that $T'>0$ everywhere and $<T'> = 1.0$.
It is simply the template obtained by normalizing the spectrum to have an average
value of unity across the wavelength of interest.
We define $\hat{t}$ and $t'$ in a similar way.
In \S~3.2, we define a norm, $|| x || \equiv \sqrt{\sum_i x_i^2} $
and ``normalized'' zero-mean template $\tilde{T} = \hat{T}/||T||$.

Strictly speaking in the treatment of photon statistics we should use
photon-weighted templates. However, over the wavelength ranges we adopt, the
photon weighting is relatively flat, and our detector is hypothetical.
Therefore in our simulations, to avoid difficulty with
end-effects and to focus on the influence of extrasolar planet
atmospheric spectral
features, we will deal with continuum normalized spectra.
In practice, we may very well need to filter out low
order terms that are irrelevant to the important aspects of the correlation
functions and which may remain for a variety of practical reasons.

The total (undispersed) signal falling at a given single spatial pixel  is,
in photons,
$S  = S_0 + s_p$.
A spectrum of this total flux may be expressed as:

$$ S_i \equiv S(\lambda_i) = S_0.T_i + \epsilon_i + s_p.t_i$$

where $\epsilon_i$ is the photon shot noise from the star at wavelength $\lambda_i$,
and the index $i$ runs 1 to $N$ for $N$ sample points or wavelengths in each spectrum,
and units are photons for $S_i$, $S_0$, $s_p$, $\epsilon_i$ with $T_i$ and $t_i$ dimensionless.
That is $S = \sum S_i$.
We presume that there is some means to estimate the spatial shape and
flux of the stellar PSF, either empirical or theoretical,
and we will attempt to subtract this estimate of the host star
PSF from the observation.
The estimate of the true flux $S_0$ will in error by a fractional amount
$\delta$ given by
$S(\hbox{estimate}) = S_0(1+\delta)$.
Then consider the difference spectrum

$$d_i = S_i - S_0(1+\delta)T_i$$
$$d_i =  \epsilon_i + s_p.t_i - S_0.\delta .T_i\eqno(1)$$

where $d_i$ is in photons,
the index $i$ corresponds to spectral dimension and, as above,
$\epsilon_i$ is the noise and units are photons.

It is the difference spectrum $d_i$ that we cross-correlate with the zero-mean
template of the planet $\hat{t}$.
Let the cross-correlation function $c_j$ be given by
$$c_j \equiv \sum_i d_i \hat{t}_{i-j} = \sum_i \epsilon_i\hat{t}_{i-j} +
s_p.\sum_i t_i\hat{t}_{i-j} - S_0.\delta .\sum_i T_i\hat{t}_{i-j} \eqno(2)$$

Each of the terms in equation (2) has a different physical meaning
and each contributes to detection capability.
In the following subsections, we look at each term individually to
understand its importance.

\subsubsection{Signal from the Planet}

The second term of Equ.~(2) represents the signal from the planet.
When the template is cross-correlated
in the wavelength dimension with the signal, the value at the zero offset
position (i.e. ``on'' the planet) from the second term is
$$c_0 = s_p \sum_{i} t_i \hat{t}_i = s_p\sum_i(\hat{t}_i + 1/N)\hat{t}_i$$
$$c_0 = s_p \sum_{i} \hat{t}_i^2\eqno(3)$$
since $\sum_i \hat{t}_i = 0$ by definition.

We will also make use of the identity $\sum_{i} \hat{t}_i^2 = \sigma^2(t')/N$,
where $\sigma(t')$ is the standard deviation of the template normalized
to have unit mean in the region of study.
This leads to $c_0 = (s_p/N)\sigma^2(t')$ for the ``signal''.
The $t'$ (unit mean) template is the most straightforward to derive
observationally, and its standard deviation offers an intuitive
description of the degree of structure in the template.

\subsubsection{Photon statistics and $S/N$}

The first term of Equ.~(2) is the photon shot noise from the host star,
and at the correlation peak, from this first term,
$c_0 = \sum_i\epsilon_i\hat{t}_i$.
If there are $S_0$ photons from the star in the pixel,
there are $S_i \approx S_0 T_i$ photons in spectral pixel $i$,
and the expected dispersion per pixel from Poisson statistics
is therefore
$\sigma_i \approx \sqrt{S_0 T_i}$.
Hence, the variance of the cross-correlation function
at peak due to shot noise from the star
is $E(c_0^2) = E((\sum_i\epsilon_i\hat{t}_i)^2)$ where $E$ denotes expected value.
It follows that $\sigma^2_{c_0} \equiv E(c_0^2) = \sum_i\hat{t}_i^2E(\epsilon_i^2)$
since the expected value of the cross terms, $E(\epsilon_i\epsilon_j)$, is zero
due to the independence of $\epsilon_i$ and $\epsilon_j$,
and hence, since $E(\epsilon_i^2) = \sigma_i^2$ by definition,
$$\sigma^2_{c_0} = \sum_i\hat{t}_i^2S_0T_i$$
$$\sigma^2_{c_0} = S_0 \sum_i(\hat{T}_i + {1\over N})\hat{t}_i^2\eqno(4)$$
(where, again, the `hat' symbol refers to a zero mean template).
The latter term, $1/N$, is independent of the shape of the stellar template
and dominates over the former, $\hat{T}_i$, for the cases we studied.

Combining equations (3) (signal) and (4) (noise), and using the identity
mentioned in the previous section,
we derive the $S/N$ of the correlation peak due
to photon statistics from the host star:
$$S/N = {s_p\over \sqrt{S_0}} {\sum_i\hat{t}^2\over \sqrt{
         \sum_i\hat{T}_i\hat{t}_i^2 + \sum_i\hat{t}_i^2/N}}$$
$$S/N = {s_p\over \sqrt{S_0}} \times \sigma \times {1\over\sqrt{1+f}}\eqno(5)$$
where $\sigma \equiv \sigma(t')$ is the standard deviation of the unit-mean
template and $f = \sum_i \hat{T}_i\hat{t}_i^2 / \sum_i \hat{t}_i^2$
which is a small quantity, for the cases we studied.

In words, the signal-to-noise $S/N$ of the peak of the correlation function for
a matched template is approximately the $S/N$ of the direct image times the
standard deviation $\sigma$ of the template in its unit-mean form.
To illustrate, the extreme situation is one where all the flux
comes from a single emission line, and the star has a flat, featureless
continuum.
Then, $\sigma = \sqrt{N}$ for the unit-mean emission-line template
and the $S/N$ is {\it increased} over direct imaging by a factor $\sqrt{N}$.
This is as expected intuitively, since we are simply excluding all flux
from the star arising from wavelengths where the extrasolar planet does not emit, equivalent
to centering a narrow filter on the emission line. The planet's flux
is preserved, but the star's, and hence contaminating noise, is decreased.
For $10^4$ spectral data points,
the $S/N$ of the planet to star is $S/N = 100 \times s_p/\sqrt S_0$:
we could see a planet a hundred times fainter by dispersing the light in this
example, if we knew that its light arose from a single emission line.
There may be practical application of this example at Ly~$\alpha$ or
other resonance scattered lines.

A single emission line maximizes the magnitude of the improvement,
however there is obviously a middle ground between a $\delta-$function
and a completely
featureless spectrum, where we may benefit
from increasing the spectral resolution.
Below, we examine plausible Jovian and terrestrial templates and find
that for typical spectral windows this standard deviation implies
a moderate {\it loss} of $S/N$ compared to a direct image.
There are however spectral windows where the parameter $\sigma$ is large,
and it increases as a function of the spectral resolution.
Even so, the spectral approach can win in the next term, which is the
elimination of systematic mis-subtraction of the host star.

\subsubsection{PSF Matching Error}

The third term in Equ.~(2),
$c_j = - S_0.\delta .\sum_i T_i\hat{t}_{i-j},$
represents
our ability to remove systematics and model or measure the
reference PSF.
This can be the limiting factor
(Brown \&\ Burrows 1990).
Again, on the correlation peak, we have the error term
$$c_0 = -S_0\delta \sum_i T_i\hat{t}_i$$
$$c_0 = -S_0\delta \sum_i \hat{T}_i\hat{t}_i\eqno(6)$$
since $\sum_i \hat{t}_i = 0$ by definition, and $T_i = \hat{T}_i + 1/N$.

Typically, if the extrasolar planet spectrum is significantly structured in a way
unrelated to that of the star, or if the velocity moves the spectrum
more than the instrumental resolution, then the term $\sum_i \hat{T}_i\hat{t}_i$
is small.
We can quantify the gain due to this systematic term as the ratio $R$
of signal-to-error, Equations~(3) and (6),
$$ R = {s_p\over S_0\delta} \times {\sum_i\hat{t}_i^2\over \sum_i \hat{T_i}\hat{t}_i}.$$

In the pure imaging case, this ratio (the planet's signal
to mis-subtracted stray light) is $R = s_p/(S_0\delta)$,
and hence using correlation techniques we gain by a factor
$\sum_i\hat{t}_i^2 / \sum_i \hat{T_i}\hat{t}_i$.
Equivalently, we reduce the host star flux by the inverse of this
factor.
This gain can be very large, from several tens to thousands.
In \S~3.2, we modify our procedure to (mathematically)
make the gain infinite.

\subsection{Gram-Schmidt Orthonormalization to Eliminate Stray Light}

Again, the noiseless spectrum at a pixel is
$ S = S_0 T + s_p t$ which is
a linear combination of the star and planet templates.
The ``basis vectors'' $T$ and $t$ are independent but not necessarily
orthogonal, and hence span a two dimensional vector space.
The correlation functions we have been dealing with provide
an inner product defined on that vector space
$$<d,t> =   \sum_i d_i t_i$$
and the norm of a function for this inner product is given by, say,
$$||<x,x>|| = \sqrt{ \sum_i x_i^2 }.$$
We can define a unit basis vector $$\tilde T \equiv {\hat{T}\over || \hat{T}||}$$
(i.e. such that $<\tilde{T},\tilde{T}> = 1$).
Similarly, it will be convenient to define a scaled version of $\hat{t}$ which
has unit norm, $\tilde{t} = \hat{t}/\sqrt{\sum_i \hat{t}_i^2}$.

Given such a space, it is always possible to generate an orthonormal set of basis vectors
that span the space, starting from any unit norm basis vector, such
as $\tilde{T}$.
(This procedure is known as Gram-Schmidt orthonormalization.)
An orthonormal basis set is one whose members {\it are} orthogonal,
i.e. their inner product is zero,
unlike the  $\tilde{t}$ and $\tilde{T}$ vectors by themselves, and they have been scaled
to have unit norm.

For a two dimensional space like ours, we can simply define the function
$$g = { \tilde{t} - <\tilde{T},\tilde{t}>\tilde{T}  \over
||   \tilde{t} - <\tilde{T},\tilde{t}>\tilde{T} ||}\eqno(7)$$

It is straightforward to show that $<g,\tilde{T}> \equiv 0$,
remembering that $\sum_i\tilde{T}_i^2 = 1$ by the definition of unit norm.

So what happens if we cross-correlate our spectra (i.e. derive the inner product)
with the Gram-Schmidt
function $g$ that may be derived for any given planetary and stellar template pair?
First, in the absence of noise, we derive our ``signal'' $<S,g> \equiv \sum S_i g_i$:
$$<S,g> = <S_0 T + s_p t, g>$$
and $T = \hat{T} + 1/N$, $t = \hat{t}+1/N$, $\tilde{t} = \hat{t}/||\hat{t}||$
$\tilde{T} = \hat{T}/||\hat{T}||$.
Hence $S = S_0(1/N + \tilde{T}\sqrt{\sum_i\hat{T}^2}) +
s_p(1/N + \tilde{t}\sqrt{\sum_i\hat{t}^2})$.
Given 
that $<g,\hbox{constant}> \equiv 0$ since we are dealing
with the zero mean versions of the templates,
and that $<\tilde{T},g> = 0$ by the construction of $g$,
it follows that the signal
$$<S,g> = s_p \sqrt{\sum_i\hat{t}_i^2} <\tilde{t},g>\eqno(8)$$
or in other words the terms involving $S_0$,
the stellar component,
and any constant offsets
{\it drop out.}
This is the same as the signal obtained from cross-correlation, but
for zero stray light error, noting
the different scalings used for the templates and modified by the inner product
of $\tilde{t}$ and $g$.

The term $<\tilde{t},g>$ measures how orthogonal the planetary
template is to the stellar template.
We can expand this inner product to find that
$$<\tilde{t},g> = (1 - <\tilde{T}.\tilde{t}>^2)^{1/2}$$
where, as in the cross-correlation section, the quantity
$<\tilde{T},\tilde{t}>$ is typically
very small.

We may, as above, derive a variance for the inner product of $S$ with
$g$ assuming only Poisson photon noise.
Similar to before,
if $w = \sum_i g_i \epsilon_i $ then $E(w^2) = \sum_i g_i^2\sigma_i^2
= S_0\sum_i g_i^2 T_i$.
Hence
$\sigma^2(w) = S_0\sum_i\hat{T}g_i^2 + S_0/N\sum_i g_i^2$, however $g$
has unit norm, $\sum_i g_i^2=1$ so

$$\sigma^2(w) =
{S_0\over N} + S_0 \sqrt{\sum_i\hat{T}^2} \sum_i\tilde{T}_i g_i^2\eqno(9)$$
which is similar to the cross-correlation version, Equ.~(4).

We can therefore write a $S/N$ for the Gram-Schmidt orthonormalization
by combining Equs.~(8) and (9).
$$S/N = {s_p\over \sqrt{S_0}} 
\times \sigma(t') \times <\tilde{t},g>/\sqrt{1+h}\eqno(10)$$
where again we have made use of the identity $\sigma(t')=\sqrt{N\sum_i\hat{t}^2_i}$
and $h = N.S_0\sqrt{\sum_i \tilde{T}^2}<\tilde{T},g^2>$.
We expect $h$ to be small, and as above, $<\tilde{t},g> \approx 1$,
so the loss in $S/N$ that might have been anticipated by moving to a truly orthogonal
basis is very small indeed.
{\it The $S/N$ of the Gram-Schmidt method is essentially the same as that
of the cross-correlation technique, but with the advantage that stray
light is completely eliminated.}

Note also that, in principle, with the Gram-Schmidt approach
we do not need to carry out additional
PSF estimation and subtraction.
In practice that may still be
desired, but under these idealized conditions it is not needed.
Finally we note that this method
will only operate effectively where the planet spectrum
differs significantly from that of the star, either because of
atmospheric spectral features or velocity shifts.
The spectral templates we constructed, below, are the as-observed Jovian spectrum
and a terrestrial template, and they do
display substantial structure which distinguishes them from the star.
Woolf et al. (2002) present the integrated spectrum of the Earth
derived from Earthshine observations, and they show empirically
that strong atmospheric absorption features are indeed prominent.

\subsection{Analysis of Variance}

Cross-correlation with a matched template offers a  good strategy
for finding a signal in a noisy dataset.
If, however, we do not have a template, progress can still be made
provided that there are spectral
differences between the star and extrasolar planet.
The noiseless spectrum at a pixel is
$S = S_0 T + s_p t$,
a linear combination of the star and planet templates multiplied by
a flux contribution from each.
The ``basis vectors'' $T$ and $t$ are independent but not necessarily
orthogonal, and hence, as discussed in \S~3.2, span a two dimensional vector space.
A variety of methods exist to determine the dimensionality
of a distribution of points in such a space.
Principal component analysis may be used to determine
how many basis vectors are needed to describe the distribution.
Closely related to this, and sufficient for our purposes
is a simple analysis of variance.

We may, in principle, derive a very high $S/N$ mean
stellar spectrum template empirically from the integral-field observation
itself, estimate the flux at each position from the observed value
and then subtract a scaled mean. Analysis of the variance of the
residuals can offer a statistical statement about the likelihood that
the photon distribution arises only from the stellar host spectrum.
The variance of the residual distribution for a given pixel
follows the $\chi^2$ distribution, and so it is easy to
quantify the probability that the observed distribution could
have arisen under the null hypothesis: that there
is no planet present.

A generalized statistical approach such as the analysis of variance is attractive because it requires
no knowledge of the
character of the planetary template.
Clearly, having established that significant excess variance exists
(or an equivalent statistical statement),
one would then interrogate the spectrum in more detail to
either derive an empirical flux distribution, or else to quantify
its similarity to a variety of predetermined templates and hence
deduce its nature.
In our simulations below, we illustrate application of a $\chi^2$
statistic.

\section{Simulations for Spatially Invariant Spectra}

Obviously these are idealized approaches in terms
of the concept of the datacube, and the statistical techniques
are straightforward.
However, they serve to emphasise the principle that there are
mathematical and statistical approaches
to data analysis, in addition to specialized hardware,which
may contribute significantly to solving problems in extrasolar planet detection
and characterization.
In \S~4 we proceed to illustrate application of the formulae of \S~3.
Given the assumption of a spatially invariant spectrum (after filtering
out low order differences), this may be considered as an idealized representation
of a time-averaged ground-based observation, or of a space based
observation over either a limited wavelength range or close to the host
star where diffraction effects do not cause high frequency ripples.

\subsection{Derivation of Template Spectra and the Photon Budget}

We will consider Solar illumination of Jupiter and the Earth as the basis
for our templates.
One of the fundamental parameters whose influence we wish to study is
spectral resolution.
We therefore attempted to obtain template spectra at the highest possible
spectral
dispersion, which may be degraded to lower spectral resolution, or not, depending
on the application.

Simulation of very high resolution spectra reflected from planetary surfaces,
particularly terrestrial ones,
is complicated by the fact that the atomic and molecular features of
most interest are the very ones that hamper astronomical observation
of astronomical targets.
Rather than strive for ultimate accuracy therefore, we have created a set of
illustrative, but realistic, templates as described in the following section.
All spectra were 
resampled onto a uniform grid
in log$\lambda$ with a constant resolution $\lambda / d\lambda = 500,000$
using standard interpolation routines.
Shifting and cross-correlation of a template onto a logarithmically sampled spectrum corresponds
to a velocity shift.

(\romannumeral1)~{\it A Solar template} was generated from the Solar model of Kurucz derived
using the program ``ASUN'' and available from Kurucz's website
{\it http://cfaku5.harvard.edu/}. The file used was {\it irrcmasun.asc} in the
[SUN.IRRADIANCE] directory. We further processed this model to normalize
by the local continuum (presented in the file)
and resampled onto a uniform grid
in $log \lambda$ with a constant resolution $\lambda / d\lambda = 500,000$.
This was done using spline interpolation.
We decided to use a model atmosphere in order to avoid problems in the
regions of most interest: namely where there is a large amount of molecular structure
arising from terrestrial absorption.
While in detail the model atmosphere may differ from the actual Solar spectrum,
it is quite adequate for our purposes in providing a statistically
realistic distribution of lines, line densities and depths.
The resulting continuum normalized template Solar spectrum is shown in Fig.~1.

(\romannumeral2)~{\it A Jovian template} was made, which is a hybrid high-resolution Solar
spectrum (the one described above) multiplied by a moderate resolution
spectrum of the Jovian albedo from Karkoschka (1994).
Karkoschka (1994) and (1998) presents and discusses the albedo of Jupiter
and the other outer planets from 3000~\AA\  to $1.050~\mu m$ .
The resolution is $\approx 10$\AA, or $R\sim 500 - 1000$.
We kept the reflected Solar component at the full resolution above.
In reality, there may well be additional fine structure apparent at high resolution
in the Jovian albedo.
For example,
Tokunaga et al. (1979) show numerous emission lines in the $10\mu m$ window
for Jupiter, emission from $C_2H_6$, and reviewed by Larson (1980).
Rotational broadening may enter at our full resolution since $R=500,000$
corresponds to velocity $v = 0.6$~km.s$^{-1}$, or 1.2~km.s$^{-1}$ for two
pixel width bin.
The
equatorial speed of Jupiter is 12.6~km.s$^{-1}$, however the majority of
the reflected light arises from the disk center regions where the relative velocity is much closer to zero, and additionally any
inclination of the planetary pole out of the sky-plane further reduces rotational
broadening.
We did not make any correction for rotational broadening in our analysis.
The resulting spectrum is shown in Fig.~2 as a function of resolution.

(\romannumeral3)~{\it A Terrestrial template} was generated with absorption features only.
Again, for an incident spectrum, we used the Kurucz model above.
To estimate the atmospheric absorption spectrum, we used the
three atlases of the Solar spectrum
Livingston \&\ Wallace (1991),  Wallace, Hinkle \&\ Livingston (1993),
Wallace, Hinkle \&\ Livingston (1998) which  present Fourier Transform
Spectra of the Sun from 3570~\AA\ to 5.4~$\mu m$ at a resolution of order 200,000
to 400,000 over the majority of the wavelength range ($\lambda > 5000$~\AA ).
These atlases present atmospheric transmission functions at high spectral resolution
which are well-suited for our purpose.
To generate a terrestrial template, we multiplied the Kurucz model by
the square of the atmospheric transmission (squared to allow for the
fact that light reflected from the surface of the Earth has to pass
through the atmosphere twice).
Fig.~3 shows the resulting terrestrial template spectrum.
In this case, as spectral resolution grows, the complexity of the
absorption features continues to grow. The dominant bands are from oxygen
and water, and may also be seen in the observed spectrum of
Woolf et al. (2002).
Fig.~4 expands the region around the prominent molecular
oxygen A-band and offers
a sense of the level of complexity that arises as spectral resolution increases.

(\romannumeral4)~{\it Another Terrestrial template} was generated with absorption
and a selected set of emission lines from the Meinel series of OH lines in the near-IR.
We were interested in understanding whether atmospheric emission lines
might be detectable from extrasolar planets.
This could only occur because they are extremely narrow, which allows,
in principle, rejection of the continuum starlight in a fashion analogous
to, but the reverse of, a terrestrial OH-suppression device.

By far the brightest line emission from the night-time airglow is
from the Meinel series of OH radical vibration-rotation complex (Meinel 1950,
Krassovsky, Shefov \&\ Yarin 1962,
Leinert et al 1998).
The intensitites are highly variable and decrease through the night
(Ramsey et al. 1992). A typical total intensity is of order 5~MR (Leinert et al 1998).
We obtained a list of bright OH lines (only) in the $1 \mu m$ --- $2 \mu m$
range in order to understand their likely relevance.
This was downloaded
from the Cambridge OH Suppression
Instrument (COHSI) web page,
{\it http://www.ast.cam.ac.uk/\~\ optics/cohsi/www/ohsky/database.htm},
which draws from  Maihara et al. (1993) and Olivia \&\ Origlia (1992).
To estimate an order of magnitude amplitude at which to insert these
lines, we first converted the tabulated intensities to Rayleighs and
then estimated an approximate terrestrial reflected Solar spectrum.
The (low resolution) flux spectrum of the Sun obtainable from the
Hubble Space Telescope (HST)
calibration library {\it crcalspec} within iraf/stsdas was used,
with a Bond albedo of 0.306 for the Earth.
The solid angle of Earth from the Sun is $5.7\times 10^{-9}$~sr,
and the total scattered flux is $1.4\times 10^{-10}$ that of the Sun,
hence the approximate surface brightness of the Earth in R~\AA$^{-1}$\ is
$R_E \approx 3.06\times 10^{-7} S_{\odot}$ where $S_{\odot}$ is the Solar
flux expressed in photons~s$^{-1}$cm$^{-2}$\AA $^{-1}$.
Given the large variability of these lines, and to better assess the difference
they make, we arbitrarily increased their intensities from the tabulated value
by a factor 5, corresponding to a total intensity of 3.2~MR in these lines.

Fig.~5 shows the resulting spectrum, again normalized to
the continuum level. As can be seen,
the OH lines do start to become evident for very high spectral resolutions,
although their strength never rises to dominate the spectra.
Additional work on atmospheric line emission further into the infrared,
or possibly ultraviolet, may prove interesting. In the IR,
some of the strongest emission lines occur where water vapour
strongly absorbs incident sunlight, and so the relative contrast between
the emission and absorption spectra may be very different.
Also, the character of the night and day emission spectra differs substanatially
(Zipf 1966, Le Texier et al 1989).

\subsection{Quantitative Characterization of the templates}

In order to design an experiment, Equ.~(5), which gives the
$S/N$ of the cross-correlation peak, may be used for a
given template to plot $\sigma$,
$s/\sqrt{S_0}$, and hence $S/N$ as a function of wavelength, for a chosen
spectral interval or resolution.
In other words, the correlation function contains more information in
regions where the spectrum is highly structured and these plots allow such
spectral windows to be identified.
Figs.~6 \&\ 7 show the case of 2\%\ sampling intervals for calculating
these parameters.
It suggests that we either go to the blue, where there are strong photospheric
absorption features from the illuminating star (in our case, the Sun), or else
to the near infrared where strong molecular absorption features in the
planet's atmosphere dominate.
Since the latter differentiate the extrasolar planet from the star, this is the
region we choose to study.

For the purposes of analysis and simulation, 
we selected a spectral region that seems to be a plausible candidate
for studies of the type we advocate.
This primary region is essentially the $I-$band, from 7000\AA\ to
1~micron, where strong water and oxygen absorption features are
present in the terrestrial spectrum, and deep, broad methane absorption
features are apparent in the Jovian spectrum.
In future work it may be of interest to select, for example,
a very narrow window around the oxygen A-band
from 7580\AA\ to 7800\AA, in order to attempt specific targetting
of terrestrial extrasolar planets with oxygen in their atmospheres, provided
the photon budget permits it.

The other parameters of the spectral
templates which offer an indication of the utility
of the correlation function are the width and contrast of the
autocorrelation function, and the variance (i.e. degree of spectral structure)
as a function of spectral resolution.
The autocorrelation functions are shown in Figs.~8 and 9, showing the very
narrow peak in the terrestrial version, indicative of the wealth of
structure
present at high spectral resolution.

Figs.~10, 11 and 12 show the run of standard
deviation of the unit-mean template with spectral resolution for
each example.
For the terrestrial template, Fig.~10, we see a $\sigma$ that increases
continuously, to the limit of our resolution. The
value of $\sigma$ is in the range 0.1 --- 0.2, corresponding to a $S/N$
about 20\%\ that obtained by direct imaging alone.
(Recall: we still have the direct image; the spectral dimension
is adding extra information.)
The A-band region shows similar behaviour, though with an
overall dispersion that is larger, corresponding to about 40\%\ of
the imaging $S/N$.
The Jovian template has a high degree of structure even at relatively low
spectral resolution,
and it saturates at resolutions of a few hundred.
This may be an artifact of the hybrid nature of the template,
or else it may be that the broad methane bands do not contain
additional fine structure.
If there is additional fine structure, then the dispersion can only
increase and hence the $S/N$ too.

Finally, we consider the ``gain'' term which is the reduction of stray
light by the cross-correlation process.
The gain term $R$ is
$$ R = {s_p\over S_0\delta} \times {\sum_i\hat{t}_i^2\over \sum_i T_i\hat{t}_i}.$$
and is dominated by the function $\sum_i \hat{T}_i \hat{t}_i$.
For the spectral regions we selected, there is little correlation between
the stellar and planetary templates.
There is no physical constraint on whether the quantity should be positive
or negative, and indeed as we shift the relative velocities, the function
can, and does, pass through zero, corresponding to infinite gain and
complete rejection of the stray stellar light (albeit not the noise from
that light).
These functions depend in a sensitive way on the template,
the spectral region, the
relative velocity and the spectral resolution, so rather than
attempt to span this large and somewhat unstable parameter space,
we hark back to \S~3.2 and note that infinite gain may be
forced mathematically.
Suffice it to say that gain values ranged from several tens to thousands
for the examples we studied.

Thus, if the dominant error term is the subtraction of the PSF,
then we could gain over direct imaging according to the product of
the dispersion of the template (which typically reduces $S/N$ somewhat) and the gain function,
which greatly decreases the contaminating starlight.

\subsection{Photon Budget}

Aside from all other considerations, it is clearly necessary that
there be sufficient photons from the extrasolar planet we seek so that
detection is at least possible.
The smallest aperture that might be considered for direct planet imaging
is $\sim 2$-m, similar to the aperture of HST.
Space telescopes of order two to ten meters diameter are under consideration
for extrasolar planet finding and characterization, while ground based
concepts can reach many tens of metres diameter.
For an order of magnitude estimate, we assume an Earth/Sun flux ratio of
$1.04\times 10^{-10}$ and a Jupiter/Sun flux ratio of $5.18\times 10^{-10}$.
Scaling from the number of photons that would be detected by
the Advanced Camera for Surveys on HST in the F814W filter (the $I$-band),
which is $2.6\times 10^{10}$~sec$^{-1}$ for the Sun at 1~pc,
we derive
Table~1 which gives the number of photons for various apertures and distances.
We also consider two bandpasses: the $I$-band and a small window centered
around the molecular oxygen A-band, a possible biomarker spectral feature.
If typical background Poisson noise levels can be brought down to
of order a few hundred or a thousand counts, there are adequate
numbers of photons from the extrasolar planets in many circumstances
that direct detection may indeed be contemplated.

\subsection{Simulation for the Cross-Correlation Approach}

Cross-correlation methods would be applicable for configurations
that result in a uniform spectrum across the field of view,
such as most ground based approaches, but also space
based either over limited wavelength ranges, using averaging
or using specialized filtering methods.
For an example configuration, we used a simulated
coronagraphic observation kindly provided by
John Krist (STScI), which is typical of those expected from
the recently installed Advanced Camera for Surveys on HST,
using its aberrated beam coronagraph. The filter used for the simulation
was the F435W filter, although since {\it we are only interested in
showing the numerical approach} this is unimportant.
The surface brightness profile of this aberrated beam PSF is similar
to estimates of the ideal performance of ground-based adaptive
optics coronagraphs
in the infrared (e.g. Itoh et al. 1998, Beuzit, Mouillet \&\ Le Mignant 2000).
We simply need a PSF that demonstrates some of the generic complexities
that might be encountered, either in space or on the ground hence
we adopt this PSF for our calculations.
Fig.~13 shows the coronagraphic image.

A second simulation was also provided with typical parameter variations seen in
HST: slight ``breathing'' differences and slight mis-centering of the star on the
two frames.
This second simulation offers us a benchmark, as it
allows us to compare the idealized mathematical results with
what is one of the most powerful actual techniques currently in use on HST, namely
``roll deconvolution''.
In this approach, a second PSF of a star is obtained with the telescope
``rolled'' around the axis pointing towards the star. The PSF remains
unchanged, while the image of the extrasolar planet is moved from
its original location.
Differencing of the two PSFs results in the roll-deconvolved image.
(This is one way of estimating PSFs, parameterized by $\delta$ in \S~3.1.)
This second simulation offers a
plausible amount of
mis-match between the two PSFs one might obtain in practice.

We are not attempting to provide the limits to which the
techniques may be pushed, so we adopt the most optimistic scenario
plausible, following Brown \&\ Burrows (1990),
which is of Jupiter orbiting
a (strictly) Solar type star at 1~pc,
at a radius of 5~a.u.,
i.e. like Alpha Cen A and B.
An reasonable upper limit to the integration time is $\sim 10^5$sec, or $\sim 30$~hr, beyond which terrestrial
planet orbital motions would begin to smear their images, so this is the fiducial integration
time we adopt.
Since the simulation is for a 2-m telescope, the equivalent 8-m observation would
correspond to a distance of $\sim 4$~pc (although hopefully such a telescope would not require
an aberrated beam coronagraph!).

There are a number of free parameters, not
least of which is the data volume consideration.
We used a $150\times 150$~pixel quadrant of the PSF, sampled 0.025~arcsec
per pixel spatially, and a spectral resolution of 800 sampled twice per
resolution element to provide spectra from 7000\AA\ to 1~$\mu$m.
This results in a datacube of dimensions $150\times 150\times 571$.

In generating the spectral dimension we assumed the PSF remained the
same over the spectral window, that the stellar spectrum was as given by
the Kurucz model template, above, and we added planets with their templates
spectra, as described above. We added Jupiter at 5~a.u., and, to test the numerical
methods, ``bright Earths'' at 1.5, 2 and 3~a.u. with 
flux ten times that of the real Earth.
The flux ratio assumed for the Jovian planet was $5.2\times 10^{-10}$ of the star
and the bright-Earths $1.0\times 10^{-9}$.
At the location of the Jovian planet, the PSF wing level is
$1.6\times 10^6$~photons per pixel, about $9\times $ the peak in the planet.
The expected ideal imaging $S/N \sim 145$ from Posson statistics, although
the empirical noise level is some $300\times $ the photon noise
due to PSF fine structure, and the planet is not visible.
For the bright Earths, the PSF wing levels are
$4.1\times 10^8, 1.6\times 10^8$ and $2.0\times 10^7$
with planet peak flux $\sim 370,000$ counts.

Firstly, to show the best that might currently be attainable
in practice, Fig.~14, we simulated
a realistic ``roll deconvolution''  by differencing the two coronagraphic simulations
from one another. Slight miscenterings and focus changes (breathing) were introduced between
the two. The Jovian extrasolar planet is visible in the top right corner of
the roll-deconvolved image.
The fine-structure noise is reduced by a factor $\sim 10$ and the planet is
visible with an $S/N\sim 6$.

We now turn to the mathematical methods of \S~3.
Fig.~15 shows an image of the cross-correlation function $c_0$ from Equ.~(2)
for the roll-deconvolved datacube
using the hybrid Jovian template;
$$c_0 \equiv \sum_i d_i \hat{t}_{i} = \sum_i \epsilon_i\hat{t}_{i} +
s_p.\sum_i t_i\hat{t}_{i} - S_0.\delta .\sum_i T_i\hat{t}_{i}$$
This time the Jovian planet is very clearly visible in the top right corner.
The $S/N$ of the simulation is close to the theoretical values given above.
We expect the photon $S/N$ to be reduced by a factor $\sigma_J = 0.39$
from the structure level in the Jovian template, to about 50---60.
The gain factor $G$ for the Jovian template in this simulation is large, at $G = 1347$,
hence we expect photon statistics to dominate.
In fact in the cross-correlation image, the peak planet flux to local (empirical)
noise gives $S/N\approx 43$ in good agreement with the theory (and much better than the
$S/N=6$ of the roll-deconvolution).

The cross-correlation with the terrestrial template, Fig.~16, also reveals the
bright Earth, but the Jovian planet does not appear (it is actually a slight
depression).
This indicates
that a good template match is important.
Sensitivity to the template match is a two-edged sword: on the one hand, detection
in the general case is made more tricky, but on the other hand,
it reveals potentially important information about the spectrum of the extrasolar planet.
The gain factor for the terrestrial template is lower, at $\approx 80$
consistent with the higher degree of residual structure seen in this image.

Fig.~17 shows the analysis of variance image, an image of reduced-$\chi^2$
with a strong peak at the location of the Jovian planet, and a second
faint one at the
position of the outer ``bright Earth''.
The probability associated with the reduced-$\chi^2$ thresholded
at $5$-$\sigma$ is shown in Fig.~18 and indeed there is a statistical signature of both planets
present.

Next, we produced images of the
Gram-Schmidt correlation function image using the Jovian template, Fig.~19,
applied to the {\it single original} (not roll-deconvolved) coronagraphic datacube. The Jovian
planet is clearly visible, and the remainder of the image shows only noise
expected from photon statistics, with numbers essentially as for
the cross-correlation ($S/N = 43$).
Fig.~20 shows the
Gram-Schmidt correlation function image using the terrestrial
template
also applied to the original coronagraphic datacube. The 
outer bright-Earth
planet is clearly visible, and there are hints of the inner
two.
If we lightly smooth this image, all three are easily seen, Fig.~21.
The remainder of the image shows only noise
expected from photon statistics. The Jovian planet does not
appear, demonstrating the importance of template matching.

A we progress through this sequence, the ability to detect extrasolar planets within
the wings of the stellar PSF increases dramatically. This is because
huge gains can be made in reducing the systematic residual term,
to the Gram-Schmidt orthonormalization limit where it is
eliminated completely.
The $S/N$ of the correlation peak is related simply to the $S/N$ of the
ideal image, multiplied by the standard deviation of the planet's
template normalized to unit mean.
If there is a large dispersion in that template, i.e. the planet has lots of spectral
structure, then we can in principle gain over direct imaging alone.
In practice, for the templates studied
here, that standard deviation was in the range $\sim 0.15 $ --- $0.8$.
Despite a loss of a factor of several in $S/N$ compared to the ideal imager,
the dramatic gains in the systematic term clearly result in a better
detection capability.
If these techniques are to be applied in practice, we must also
understand the degree to which the assumption of spatially similar
spectra can be realised, a study which has many free parameters and
which is beyond the scope of this work.

\section{Diffraction Dominated PSF: Spectral Deconvolution}

\subsection{Concept and Illustration}

While the previous section explored the concept of pattern recognition
where the spectra of each pixel were identical, we now look at the effects
of potentially strong amplitude modulation of spectra introduced by wavelength
dependencies in the PSF.
A monochromatic diffraction limited coronagraphic PSF with a plausible suite of aberrations,
contains a spectacular degree
of structure.
This may broadly be characterized as a set of rings, the Airy pattern
due to diffraction by the telescope optics, with spacing
at the diffraction limit of the telescope $\lambda/D$ (where D is to be corrected
by the Lyot stop)
together with a multitude of speckles
due to scattering from mid-frequency surface irregularities
also of typical dimensions $\lambda/D$.
As wavelength increases, the ring pattern scales to larger size proportional to
the wavelength, and the speckles also grow larger proportional to wavelength.
At an angular radius $x$, therefore, we expect a characteristic change
of the spatial PSF (with wavelength) to produce a modulation
(i.e. variation in apparent flux as Airy rings and speckles move across the pixel)
in the spectral
dimension at a characteristic scale of
$dln\lambda\equiv d\lambda/\lambda \sim (\lambda/D) / x$ where $x$ is in
radians, i.e. (spatial resolution )/(angular separation).
For $\lambda \approx 1$~$\mu m$ and $x \sim 5$~arcsec,
$d\lambda \sim 200$\AA . In other words we expect modulation on a characteristic
spectral scale of $d\lambda/\lambda \sim 0.02$.

It is possible to remove some of this spectral modulation
using a variety of techniques.
Ftaclas et al (1994), for example, advocate use of very broad bandwidth
imaging to suppress the fine structure.
Fourier filtering could be used (since there is
a very characteristic set of spectral frequencies involved), however
in the example used above, the spectral frequencies of the hybrid Jovian
template are quite similar to those of the diffraction modulation,
and so Fourier filtering runs the risk of eliminating much of the
signal from the planet.

Instead, making a virtue of necessity,
we offer a simple and rather effective method to eliminate
the stellar PSF in this case, which we call ``spectral deconvolution''.
To illustrate, we again used the code of John Krist to generate a
set of monochromatic PSFs from 7000\AA\ to 1$\mu$m for a 2.0-m telescope
with a Lyot-coronagraph reducing the pupil diameter by 20\%.
A Gaussian occulting spot with HWHM = 0.3~arcsec was used.
A mid-frequency rms surface error of 0.5~nm was assumed, significantly
improved over that of HST.
A suppression factor for the host star of $0.6\times 10^{-3}$ results.
The PSF was sampled
with 0.025~arcsec pixels and a field of view $300\times 300$ pixels (10.6~arcsec
corner to corner) and again a spectral resolution of 800 corresponding to 571
spectral slices uniformly spaced in log$\lambda$.
To add planets we generated a simulated self-consistent PSF
for a point source not lying beneath the coronagraphic hole.
We again used Jupiter at 5~a.u. and extra planets at 1.5, 2 and 3~a.u.
with ten times Earth flux.
The flux ratio of the Jovian planet and bright Earth's to the host star was
$5.6\times 10^{-10}$ and $1.0\times 10^{-9}$ respectively.
This time we did not impose spectral structure on either the star or planets,
but simply adopted the same total counts in each spectral slice.
In principal spectral realism could be included, however the simplicity
of the concept proposed is quite well-shown without this additional complication.

The extrasolar planetary system was placed at a distance of 2~pc,
and fluxes were normalized
to those of a Solar type star at that distance for a 2-m telescope and $10^5$sec
integration time.
Poisson noise was added to each image.

Fig.~22 shows the average PSF of the datacube, with the extrasolar planets.
Clearly the planets are completely hidden in the scattering wings of the
stellar PSF.
Following Brown \&\ Burrows (1990), the structure is dominated by
PSF fine structure. The empirically measured noise at the location
of the Jovian planet is $\approx 75\times$ the photon shot noise.
The amplitude of the stray light in the wings of the stellar PSF is
$\approx 8.9\times 10^5$~photons per pixel compared to a peak flux
in the Jovian planet of $11,800$~photons per pixel.
That is, the planet's peak flux is only 1.3\%\ that of the local
stray light. In the notation of Brown \&\ Burrows (1990), $Q=0.013$.
If we were photon noise dominated, the $S/N \approx 12.5$ for such an
observation, however the PSF fine structure results in $S/N \sim 0.2$
and the extrasolar planet is undetectable.

The four left hand panels of Fig.~23
show slices through the datacube prior to the light smoothing that
was applied, at four separate wavelengths from
7000\AA\ to $1\mu$m (bottom to top).
Inspection of these figures shows the expected expansion of the PSF with wavelength
as features move out and everything scales with wavelength.
Fig.~24 picks a pixel at the radius of the Jovian planet, and shows the spectrum
along the datacube at that pixel.
The modulation at this resolution is very large due to these diffraction effects
as structures move across the pixel.

Now we adjust the spatial scale of the datacube as we progress through
it in wavelength. An image at 9000\AA , say, is shrunk by a factor 9/7 to match the
PSF at 7000\AA\  and a new datacube is constructed.
The four right hand panels of
Fig.~23 in contrast to the previous datacube show that there
is far
less difference between the PSFs as a function of wavelength when resampled spatially.
A plot along the spectral dimension at the same pixel as before, Fig.~25,
now shows only a smooth,
slowly varying component that can easily be modelled with a polynomial fit.
The procedure is to make such a fit, subtract it from the datacube, and then
resample the data back onto the original spatial scales.

If there is additionally a smooth scattering component as well as the diffraction
dominated speckles and rings, that too would enter as a low spatial frequency
component and would not make any difference to the effectiveness of the method.

In the spatially resampled datacube, apart from the first few slices, {\it the planet
has been moved away from its original pixel}.
It now occupies a straight diagonal slice through
the datacube starting at its original location (radius $r_0$) and ending at a radius
$r_1 = (\lambda_0/\lambda_1)r_0$ along the radius vector to the planet from the star.
The fractional width along the spectrum is $dln\lambda/\Delta ln\lambda \sim dx/(r_0 ln(\lambda_1/\lambda_0))$
where $\Delta ln\lambda$ is the logarithmic width of the entire spectral window,
and $dx$
is the diffraction width of the planet's PSF. Note that the shrinking of images
with increasing wavelength preserves this width.
{\it Hence, when a low order polynomial is fitted to the spectrum at a particular pixel,
the flux from the extrasolar planet is largely ignored}, as it represents a high order term
that the polynomial cannot reproduce.

Having subtracted the emission from the star, we can now reconsitute
the original (but subtracted) datacube by resampling the
subtracted data back onto its original spatial scale.
When the datacube is resampled back to its original dimensions and scale,
this residual planet flux realigns itself once more, and the elements of
its flux line up spatially.
Collapsing the datacube
along the spectral dimension makes the extrasolar planet become visible.

In addition, while the host star spectrum at the location of the extrasolar
planet is very strongly affected by diffraction, {\it the spectrum of the extrasolar planet is unaffected}. The planet location is obviously the center of its Airy
pattern and hence, apart from a slow change in peak brightness as resolution
decreases to longer wavelength, the spectrum of the extrasolar planet is
fully preserved.

Fig.~26 shows  exactly such an image from the simulation described.
Light spatial smoothing was first applied to avoid sampling problems
(a 2-D Gaussian $\sigma=1$).
The NOAO iraf utility {\it imlintran} was used with `sinc' interpolation to
resample the images, both to compress and re-expand, and a
fourth degree polynomial fit and subtraction was carried out for each
``spectrum''.
The final subtracted, reconstructed datacube was collapsed along the spectral
dimension to produce the image shown in Fig.~26.

The Jovian planet is easily visible, as is the outer one of the additional
planets we added. Interior to this (3~a.u. at 2~pc), residuals from the star again
dominate, and the effectiveness of the method diminishes as the ratio $dx/r_0$ increases
and the planetary signal becomes lower (Fourier)
frequency in the spectral dimension.
Quantitatively, the peak in the Jovian planet's signal is $\approx 6000$~photons
per pixel and the empirically measured noise around it is $\approx 290$~photons per
pixel. Given that we spatially smoothed a noise field which originally had
$\sigma = 950$~photons per pixel we expect the Poisson noise to reduce to
$\sigma \approx 220$~photons per pixel, consistent with our measurement.
This smoothing also accounts partially for the slightly lower peak planetary flux.
In other words, the detection achieved is $S/N \approx 20$, compared to a Poisson
limit in the range 12---40 depending on the level of spatial smoothing employed.
Recall that the peak flux of the extrasolar planet is only 1\%\ of the level
of the PSF wings.

In this simulation $S_0 \sim 9\times 10^5$,
$N\sim 600$ so the requirement that the noise not be
dominated by detector readnoise is $\sigma_R < 38/\sqrt{n_{exp}}$.
For a readnoise of, say, 4~photons, we could take 100~exposures
and satisfy this constraint.
Alternatively we may reduce the spectral resolution and shorten
the exposure time of individual exposures, provided we
keep the extrasolar planet signal in only a small fraction of
the spectrum in the resampled datacube.

Fig.~27 illustrates the idea further by showing cuts through the
datacubes along one spatial and one spectral dimension, for both
the original cube and the spatially resampled cube.
The diverging difraction pattern changes to a series of
smoothly changing ``spectra'' that can be modelled and subtrated, leaving the
planet spectrum as a diagonal in that datacube.
Improved tuning of the reconstructed image may be obtained
by allowing for the varying noise as the stray light amplitude
varies with wavelength, with $1/\sigma^2$ weighting being optimal.

This simulation was carried out in a very simple and straightforward fashion.
It ought to be possible to reach the ideal Poisson imaging limit
with this method.
The factor that would limit the derived $S/N$ is the level to which a spectrum
can be modelled and subtracted with low frequency components. If
there were high spectral frequency features {\it arising from the diffraction and scattering}
(not intrinsic to the source),
they would enter into the noise of the final image.
The expected frequencies are low, however, and no high frequency features were
found in the simulation we ran.
In practice of course the star has a somewhat complex spectrum,
however it is a well-known spectrum and it should be possible to adjust its
low-order shape to achieve a good match to the observations.
Ultimately, it may even be possible to combine a matched filtering or orthonormalization
approach with spectral deconvolution
method (e.g. by testing whether a specific pixel hosts
an extrasolar planet using corresponding matched filters in the appropriate adjacent pixels
of the spatially resampled datacube).

If we scale the simulation to an optical coronagraphic 8-m TPF concept, with mid-frequency
surface errors unchanged (0.5~nm), then the radial distance of the planet
would be now 1.2~a.u. and the background and planet would have
$\sim 16\times $ the flux. Consequently, to detect a terrestrial luminosity
planet five times fainter than Jupiter to $5$-$\sigma$ we would require
$\sim 25,000$~sec or 7~hrs.
If we allow spatial smoothing to improve S/N, as
before, the required exposure time is only $\sim 10^4$~sec.
Equivalently, Jovian planet detection for a 2-m telescope requires only
$\sim 6,000$~sec, scaling the $20$-$\sigma$ result in $10^5$~sec back to
$5$-$\sigma$.

There are a number of very appealing aspects to this method. Firstly, we can essentially
achieve the theoretical Poisson limit for direct imaging, irrespective
of the spectral character of the extrasolar planet. By contrast
the $S/N$ achievable in the cross-correlation
and Gram-Schmidt method
depends on the variance of the planet's spectrum.
Standard coronagraphic techniques are typically limited to {\it much} higher
flux levels due to the mid-frequency errors.
Secondly, and obviously, having located the extrasolar planets using a spectrum,
we have at
our disposal a spectrum of the extrasolar planet from the outset and so we can in a very
general way immediately begin to study its character.
Thirdly, we only require a single observation to detect the
candidate planet, hence we need PSF stability only over the duration of the observation.
Since we design the experiment to be Poisson
limited, the shortest possible integration times could be used
and processed independently.
This may alleviate PSF stability issues that are of concern under some circumstances.

\subsection{Related Observations and Applications}

A variation on the theme of this method is to obtain multiple
images using narrow-band filters sequentially, and then
pursue similar resampling and subtraction (spectral deconvolution)
with a datcube built from the sequential filter observations.
The wavelength sampling interval needs to satisfy two constraints
(\romannumeral1)~the planet should move by a significant distance
when the image scale is adjusted in proportion to wavelength
(\romannumeral2)~there needs to be enough data points in
the pseudo-spectrum that the planets spectrum occupies only a small
element within it.
(An equivalent strategy is to build a datacube using a long-slit spectrograph
with adjacent spectra taken sequenatially, as in Fig.~27.)

Recall from the above discussion, that to move a full diffraction
width of $dx = \lambda / D$, we would require $\Delta \lambda = \lambda \times dx/x$.
This approach offers a straightforward route to experimental
design.
For example, if $dx = 0.05$~arcsec and $x = 2$~arcsec, at 5000\AA\
we would need to change wavelength by 125\AA , which is well-suited
for example
to the parameters of the narrow-band ACS ramp filters.
Clearly it would be highly desirable to obtain more
realistic simulations, together with some exploratory observations
to determine how well this works in practice in the
presence of slowly changing optics.

In principle there are three major advantages and one disadvantage
compared to roll-deconvolution.
(\romannumeral1)~The telescope does not need to be moved. We eliminate
aberrations  due to mis-centerings between successive
observations that plague roll-deconvolutions.
(\romannumeral2)~We are insensitive to low frequency modulation which
telescope ``breathing'' would typically cause, and we can reduce this
further by keeping exposures short.
(iii) We are always on-target so the planet flux continues to
integrate. There is no dead time while reference PSFs are obtained.
On the other hand, the disadvantage is that we no longer ``spectrally
multiplex'' and so additional exposure time is needed to accumulate
the same number of photons.
If we are only interested in detection, rather than characterization,
then we would need to consider observations in the blue where spatial resolution
is improved, versus the red where the speckle pattern becomes less severe.

The spectral deconvolution technique, if it lives up to its
promise, has the potential to be applied to a high fraction
of space-based coronagraphic observations, beyond planet searches.
These may include probes of dust disks around stars and the host
galaxies of QSOs.
The case of QSOs may be particularly interesting as the combination of
strong emission lines in the QSO and possibly its surroundings
will provide a challenge to data analysis, and raises the
possiblity of trying to combine correlation and spatial scaling
techniques.

This method depends on the datacube possessing mostly
diffraction behavior (a smooth scattering component would not reduce
its efficacy), and hence is most appropriate for space-based concepts.
If the bandwidth is small (as might be the case
for focussing on specific spectral features)
or if the PSF structure is color-insensitive 
(e.g. as caused by the atmosphere)
then the cross-correlation approaches above
are likely to prove more effective.

\subsection{Practicalities}

Application of any of these ideas to real optical systems will 
require careful consideration of numerous practical
difficulties.
Not least of these is construction of an integral field spectrograph of
the dimensions envisaged, especially in the first part of
the paper.
Inefficiencies in the optical system inevitably arise in
a real spectrograph, and difficulties may arise from
scattered light internal to
the spectrograph and interference fringing
within the layers of the detector itself.
Many CCDs are especially prone to moderate or large amplitude
fringing in the very red regions in which we are most interested.

Data volume, too, is likely to be an issue.
Ideally one would like a large spatial coverage, high spatial resolution,
and moderately high spectral resolution.
The reason this problem does not need to become completely
uncontrolled is
that we seek planets close to their host stars, and hence the spatial coverage
may be restricted to the region of interest.

Finally, if compromises must be made such as abandoning
spectral multiplexing, then the degree of achievable stability enters
the trade space as we design our observations.

Despite these and other difficulties that may be encountered in practice,
it will be worthwhile to obtain a set of test observations and gain
actual experience of dealing with the issues,
and to assess the degree to which we may really approach
the limiting, idealized results.

\section{More on Planetary Spectra}

The cross-correlation techniques depend critically on the nature of the
planet's spectrum for their effectiveness.
We consider therefore a number of aspects of planet spectra that bear on
this but which may eventually lead to important new
scientific information on the nature of the planet as one might go
to higher spectral resolution with future very large aperture
planet characterization missions or ground based projects.

All other things being equal, it is best to use the highest spectral
resolution possible for a given bandpass constraint.
The degree of structure, spectral variance, and of course scientific information
such as relative line strengths always increases as spectral
resolution increases.
Even though the signal from the planet can be quite low, that of the star is typically high,
and so we do not lose to detector readnoise as we would in a more conventional
high resolution observation.
As above, \S~2, detector readnoise is one of the parameters entering concept
designs and viable observing strategies.

As higher spectral resolutions are used, cross-correlation with a matched template
more precisely locates the velocity of the planet.
Monitoring of line locations
allows not only discrimination against chromospheric activity on the host star
but vital information on the orbit of the planet, and in turn
on the mass of the planet.
At the highest spectral resolutions, planetary emission features may become
visible from night and day-glow lines as well as aurorae.
If we can  get into this regime, a new suite of diagnostics becomes
available and detection can be optimised as lines would typically be extremely narrow,
in contrast to the underlying relatively smooth stellar continuum.

The width of the correlation function (or equivalently absorption
or emission line features
in the planet's atmosphere) contains additional information,
from random thermal velocities as temperature indicators,
to systematic line broadening due to the spin of the planet in its daily cycle.

As with most other time series observations of planets, we would expect
to probe
periodicities in intensities
induced by the planets diurnal cycle.
Ford, Seager and Turner (2001) discuss large albedo changes that
may be recognizable in photometric observations of planets.
If both the spin rate of the planet  and the length of the
day can be found, then we may estimate the geometric size of the
planet.

On a longer term, yearly seasonal changes may be sought and changes due
to ``solar'' cycles on even longer timescales.
Atmospheric emission lines are especially sensitive to many changing conditions,
such as local weather or surface conditions
and atmospheric instabilities.
The terrestrial OH Meinel system in the near-IR,
for example, is intimately related to the chemistry of the terrestrial
ozone layer and changes completely between day and night
(Zipf 1966, Le Texier et al 1989).
Other atmospheric emission lines arise from oxygen, atomic
and molecular, sodium and nitrogen and are, to varying degrees,
diagnositics of the presence of life on Earth.

\section{Conclusions}

We have considered  the use of integral-field spectroscopy together with coronagraphy
as an interesting approach to planet {\it detection} as well as characterization.
The concept is to acquire a datacube of images, each at a different wavelength,
of a star with planets shining by reflected light of the star
modified by the integrated albedo (or emission) of the planet.
The star is presumed to be occulted by the coronagraph, however there remains
significant leakage or stray light from the star that dominates the problem of
finding extrasolar planets.
A potentially effective compromise modification to an integral-field spectrograph
would be construction of a spectral datacube using narrow-band filters sequentially
or a long-slit moved perpendicular to its length.

In the limiting case where the spectrum of the host star is the same across the field
of view (or a region of interest within the field
of view),
we may use pattern recognition techniques to
find extrasolar planets within the noisy spectra of the field.
This is effectively done with cross-correlation using
matched templates that in turn
offer scientific insight into the nature of the spectrum of the planet.
The $S/N$ of the cross-correlation approach is the $S/N$ of the
ideal imaging case mutiplied by the standard deviation of the planet's spectrum
(normalized to unit mean), which leads to pressure to use higher spectral
resolution.
Substantial gains can be made in rejecting the systematic contamination
from the host star.
In this limiting case, we may also use a general analysis of variance approach to
estimate whether there is any source within the field other than
just the host star, irrespective of its spectrum.

A method is described which mathematically completely eliminates
the parasitic emission from the host star, although not the Poisson noise
of that emission.
This technique uses
Gram-Schmidt orthonormalization, with the stellar spectrum and planet's template providing
basis vectors from which orthonormal vectors can be constructed.
The $S/N$ of the Gram-Schmidt method is essentially the same as that
of the cross-correlation technique, but with the advantage that stray
light is {\it completely} eliminated.

In the case where the optics are diffraction limited, although
not necessarily perfect, we describe a quite different technique to remove the
stray light from the star.
We call this ``spectral deconvolution'', applicable
to space-based observation and perhaps some ground-based adaptive optics
approaches.
In the diffraction limit, the spatial scales of the parasitic emission
are proportional to wavelength.
By numerically adjusting the image scales in the datacube to
that of a reference wavelength, the PSF changes with wavelength are much less
dramatic.
We can fit a low order spectral function at each pixel of the resampled (``shrunk'')
datacube and subtract it to approach the Poisson noise limit.
However since the planet now occupies many spatial pixels and only a limited spectral
window in the resampled datacube, its contribution to the flux is overlooked
by the fitting procedure. When the star-subtracted datacube is reconstructed
back onto the original scale, the planet spectrum is realigned and
the planet may be seen by collapsing the datacube in the spectral dimension
to make an image.
This method offers, with a single observation, essentially {\it Poisson
limited detection capability.}
At the same time, and as part of the discovery
observation, it provides a spectrum of the
extrasolar planet from the outset and suitable
for scientific investigation into
the composition of the extrasolar planet's atmosphere and possible
biomarkers.

The parameter space encompassed by integral-field spectroscopy with coronagraphic
imaging is vast, covering spatial extent, spatial resolution, spectral coverage,
spectral resolution for telescopes of various sizes both ground
and space based.
The technical and practical difficulties are non-trivial, and
there is a great deal of work required to understand which
regimes may be most effectively pursued
with techniques similar to those described here, or with much more sophisticated
pattern recognition techniques, yet the promise seems sufficiently
high that we feel the effort is likely to be worthwhile.
If these methods do prove practical,
the design of future instruments, and perhaps even
missions optimized for planet detection and characterization, may
be influenced, since it not only improves detection capability, but
at the same time and as part of the discovery observation itself,
can offer vital information on the character of
the planet's spectrum,
including the presence
of biomarkers and evidence for life.

\acknowledgements

We thank Prof. R. Kurucz for information and discussion relating to the use
of his Solar models. We also thank Prof. Lloyd Wallace for similar discussion
concerning the empirical Solar atlases.
We are exceedingly grateful to John Krist for his assistance in
producing the simulated coronagraphic images, and to Mark Clampin and
Rick White for insightful
discussion of the techniques described.

\clearpage

Bacon, R., Adam, G., Baranne, A., Court\`{e}s, G., Dubet, D.,
Dubois, J.P., Emsellem, E., Ferruit, P., Georgelin, Y., Monnet, G., P\'{e}contal, E.,
Rousset, A., Say\`{e}de, F., 1995, A\& AS, 113, 347.

Beuzit, J.L., Mouillet, D., Le Mignant, D., 2000, ESO documents
GEN-MAN-VLT-11620-1834, ``ADONIS Coronograph User Manual''

Boccaletti, A., Moutou, C., Labeyrie, A., Kohler, D., Vakili, F., 1998, A\& AS, 133, 395.

Brown, T., 2001, ApJ, 553, 1006.

Brown, R.A., Burrows, C.J., 1990, Icarus, 87, 484.

Charboneau, D., Brown, T.M., Noyes, R.W., Gilliland, R., 2002, ApJ, 568, 377.

Collier Cameron, A., Horne, K., Penny, A., Dick, J., 1999, Nature, 402, 751.

Collier Cameron, A., Horne, K., Penny, A., Leigh, C., 2002, MNRAS, 330,187.

Des Marais, D.J., Harwit, M., Jucks, K., Kasting, J., Lunine, J.,
Lin, D., Seager, S., Schneider, J., Traub, W., Woolf, N.,
JPL Publication 01-008, ``Biosignatures and Planetary Properties
to be Investigated by the TPF Mission''

Ford, E.B., Seager, S., Turner, E.L., 2001, Nature, 412, 885.

Ftaclas, C., Nonnenmacher, A.L., Grusczak, A.,
Terrile, R.J., Pravdo, S.H., 1994, SPIE, 2198, 1324.

Itoh, Y., Takato, N., Takami, H., Tamura, M., 1998, PASJ, 50,55.

Karkoschka, E., 1994, Icarus, 111, 174.

Karkoschka, E., 1998, Icarus, 133, 134.

Kasting, J.F., 1996, Ap\& SS, 241, 3.

Krassovsky, V.I., Shefov, N.N., Yarin, V.I., 1962, Planet Space Sci. 9, 883.

Kuchner, M.J., Traub, W.A., 2002, ApJ, May 2002; astro-ph/0203455.

Larson, H.P., 1980, ARAA, 18, 43.

Leinert, Ch., Bowyer, S., Haikala, L.K., Hanner, M.S., Hauser, M.G., et al.
1998, A\& ASupp, 127, 1.

Le Texier, H., Solomon, S., Thomas, R.J., Garcia, R.R., 1989, Annales Geophysicae, 7, 365.

Livingston, W., Wallace, L., 1991, NSO Technical Report \#91-001,
Tucson: National Solar Observatory,
``An Atlas of the Solar Spectrum in the Infrared from
1850 to 9000~cm$^{-1}$''

Maihara, T., Iwamuro, F., Yamashita, T., Hall, D.N.B., Cowie, L.L.,
Tokunaga, A.T., Pickles, A., 1993, PASP, 105, 940.

Marcy, G.W., Butler, R.P., 1998, ARAA, 36, 57.

Marois, C., Doyon, R., Racine, R., Nadeau, D., 2000, PASP, 112, 91.

Meinel, A.B., 1950, ApJ, 111, 207.

Olivia, E., Origlia, L., 1992, A\& A, 254, 466.

Racine, R., Walker, G.A.H., Nadeau, D., Doyon, R., Marois, C.,
1999, PASP, 111, 587.

Ramsey, S.K., Mountain, C.M., Geballe, T.R., 1992, MNRAS, 259, 751.

Roddier, F., Roddier, C., 1997, PASP, 109, 815.

Rouan, D., Riaud, P., Boccaletti, A., Cl\'{e}net, Y., Labeyrie, A., 2000, PASP, 112, 1479.

Schindler, T.L., Trent, L., Kasting, J.F., 2000, Icarus, 145, 262.

Spergel, D., Kasdin, J. 2001, AAS, 199, 8603.

Tokunaga, A.T., Knacke, R.F., Ridgway, S.T., Wallace, L., 1979, ApJ, 232, 603.

Tonry, J., Davis, M., 1979, AJ, 84, 1511.

Wallace, L., Hinkle, K., Livingston, W., 1993, NSO Technical Report \#93-001,
Tucson: National Solar Observatory,
``An Atlas of the Photospheric Spectrum from 8900 to 13600~cm$^{-1}$''

Wallace, L., Hinkle, K., Livingston, W., 1998, NSO Technical Report \#98-001,
Tucson: National Solar Observatory,
``An Atlas of the Spectrum of the Solar Photosphere from 13,500 to
28,000~cm$^{-1}$''

Woolf, N., Angel, J.R., 1998, ARAA, 36, 507.

Woolf, N.J., Smith, P.S., Traub, W.A., Jucks, K.W., 2002, ApJ, in press;
astro-ph/0203465

Zipf, E.C., 1966, J. Geomag. and Geoelec, 18, 301.

\clearpage


%

\begin{figure}
\plotone{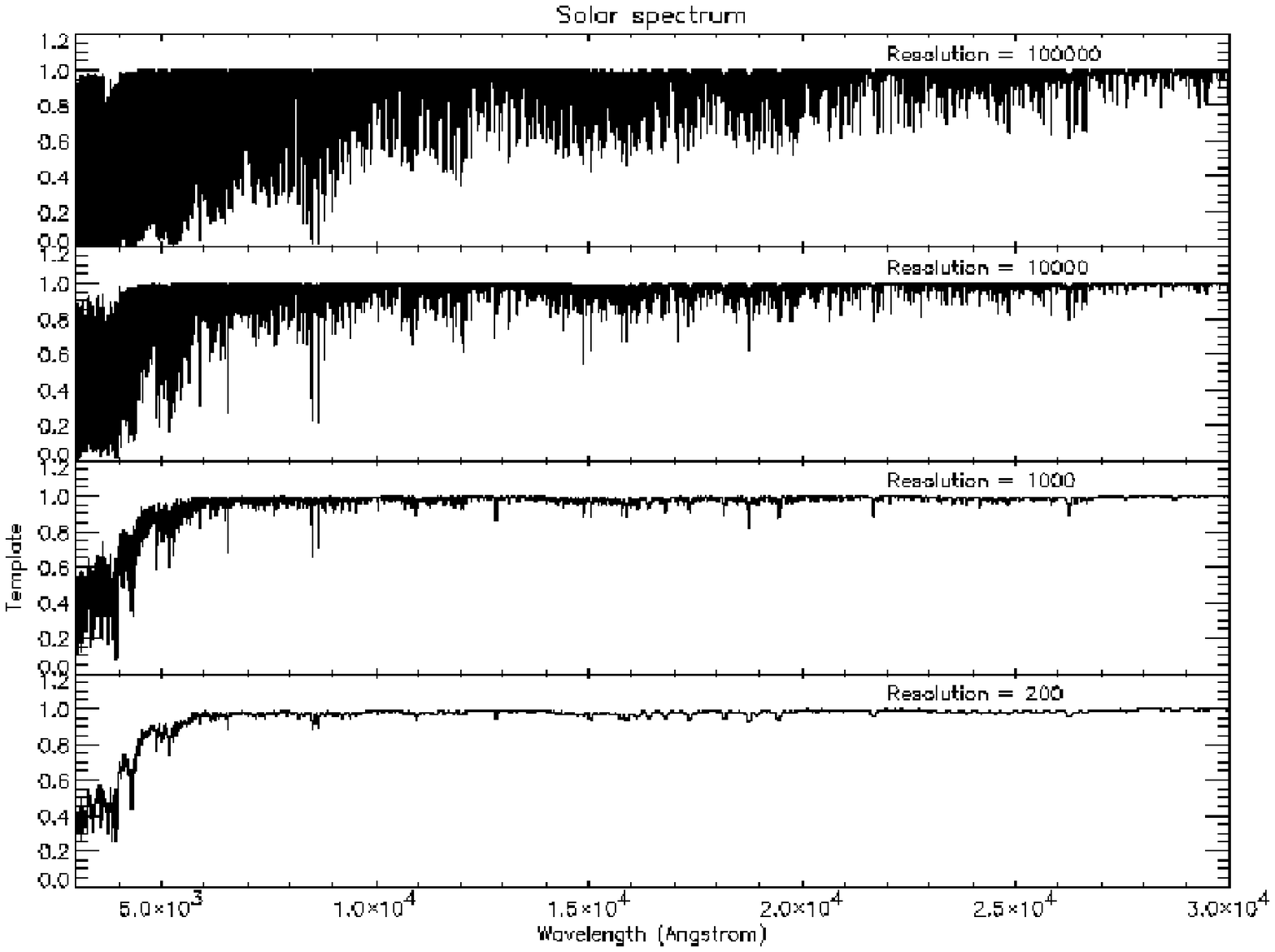}
\end{figure}

\clearpage

\begin{figure}
\plotone{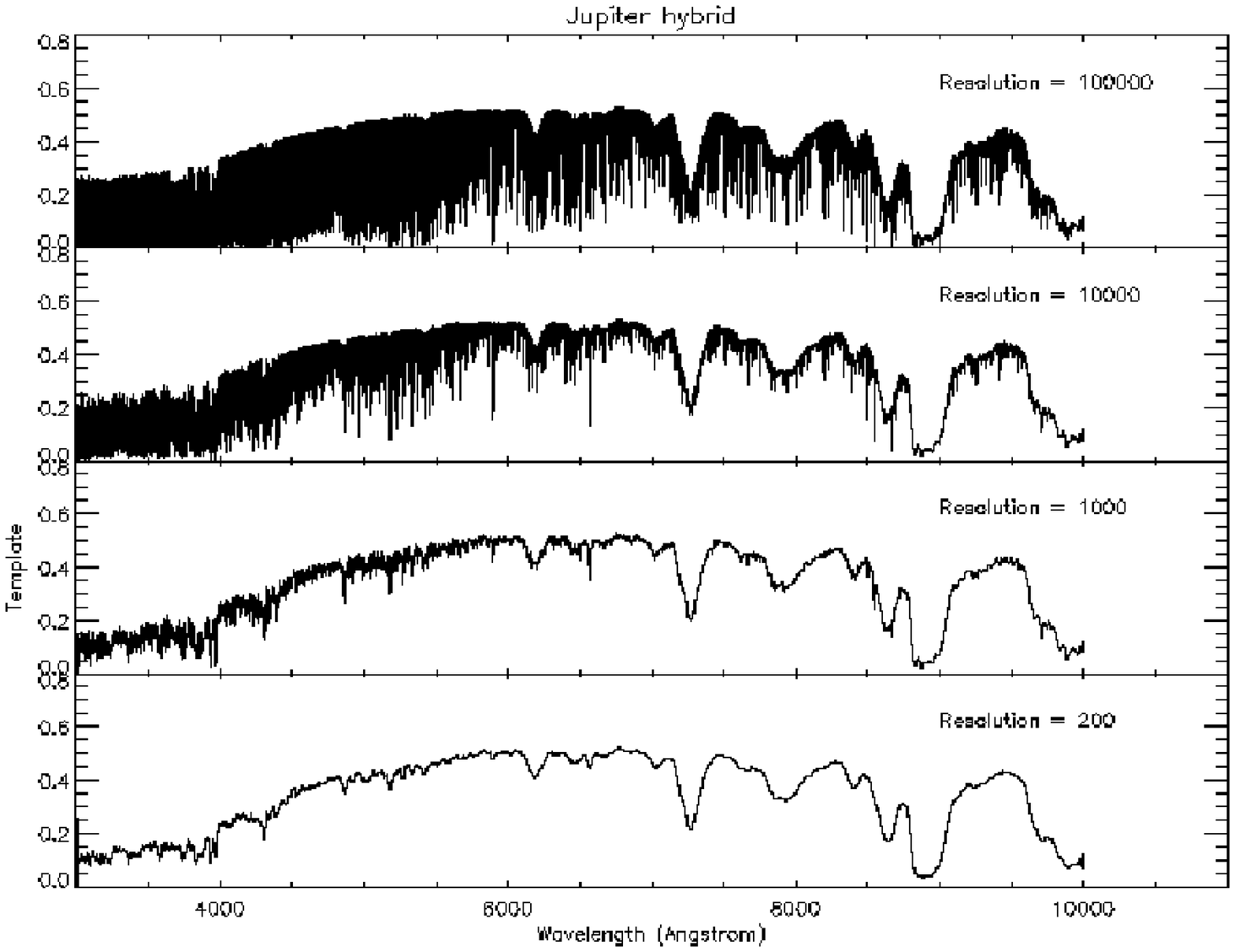}
\end{figure}

\clearpage

\begin{figure}
\plotone{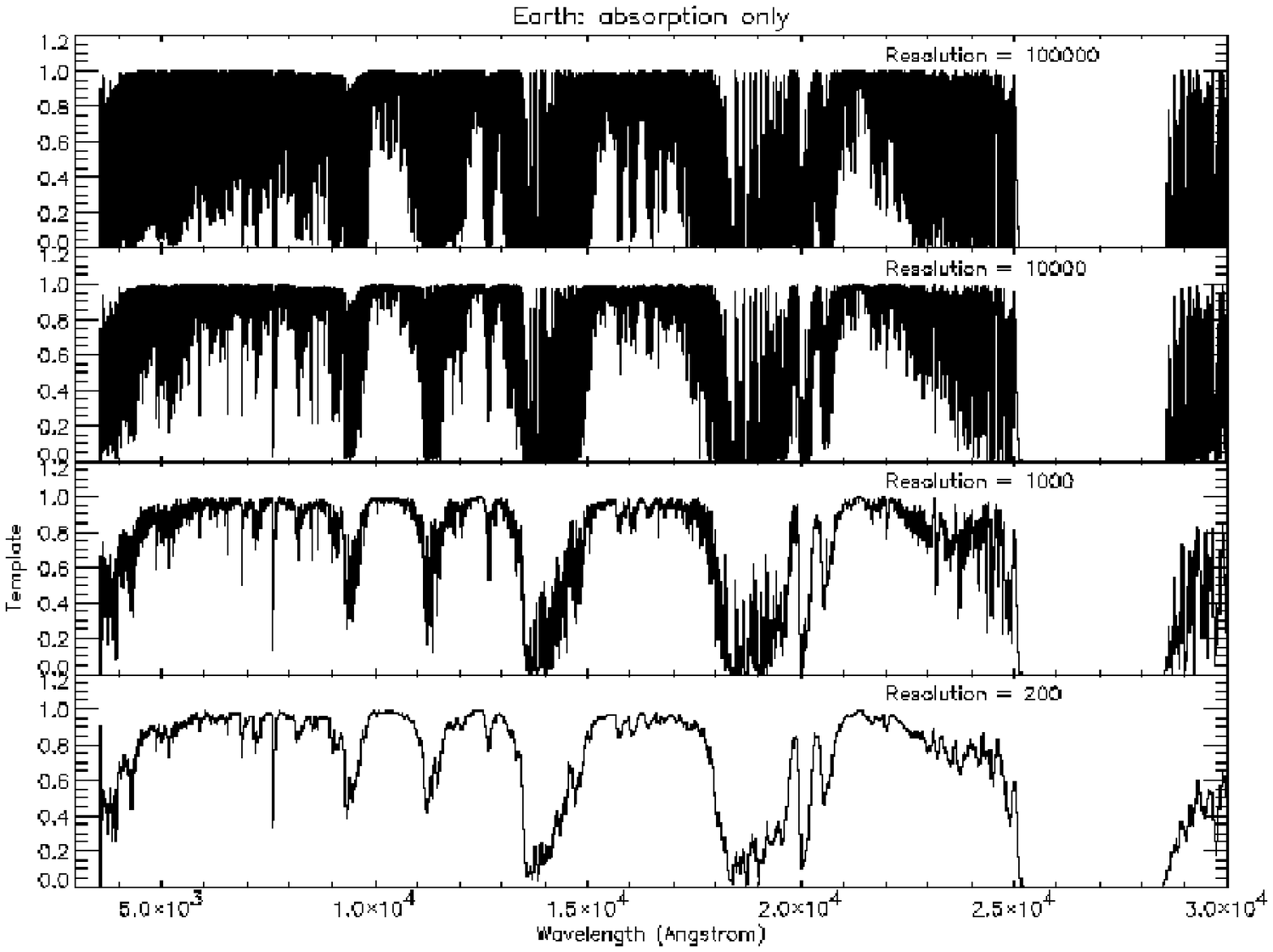}
\end{figure}

\clearpage

\begin{figure}
\plotone{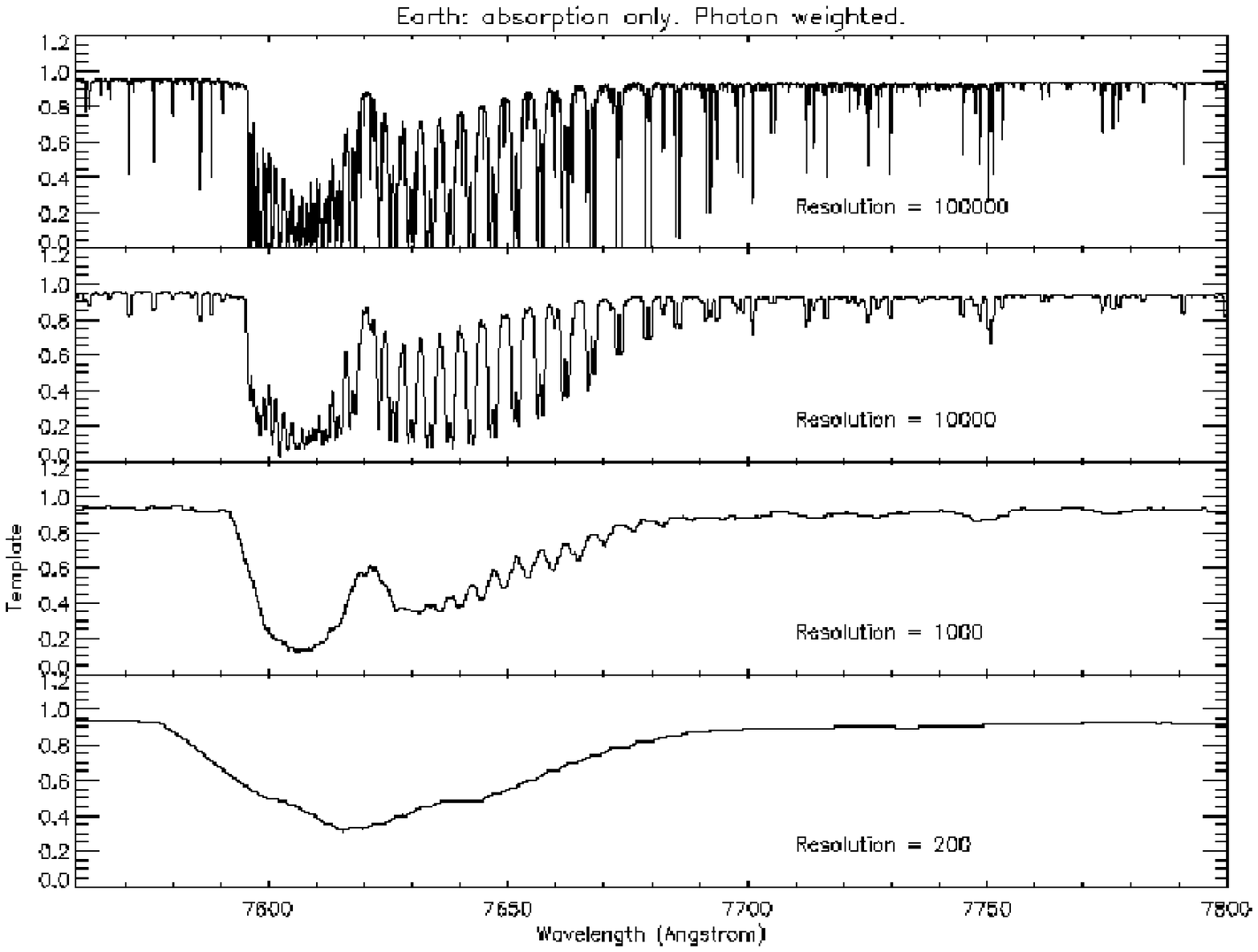}
\end{figure}

\clearpage

\begin{figure}
\plotone{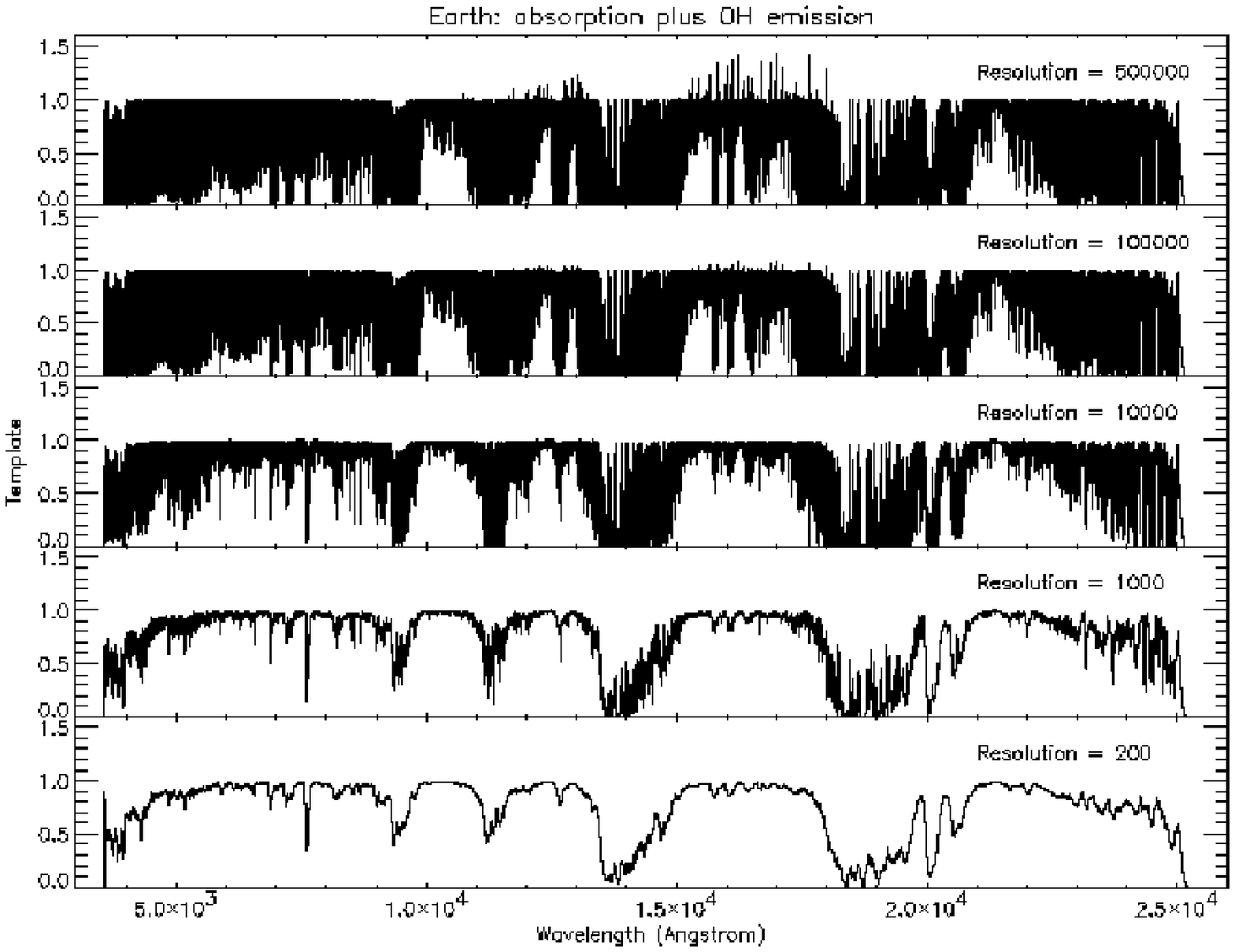}
\end{figure}

\clearpage

\begin{figure}
\plotone{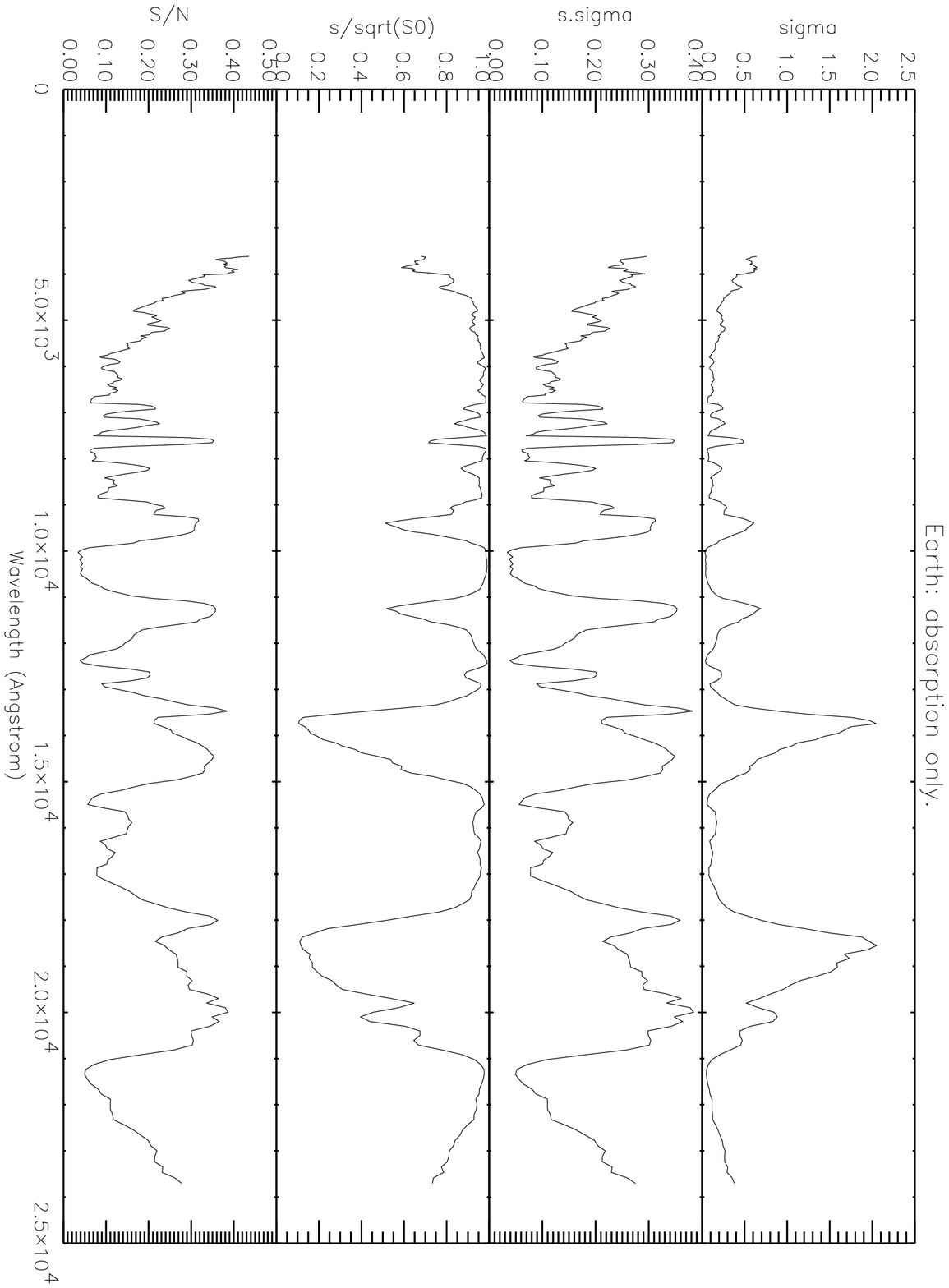}
\end{figure}

\clearpage

\begin{figure}
\plotone{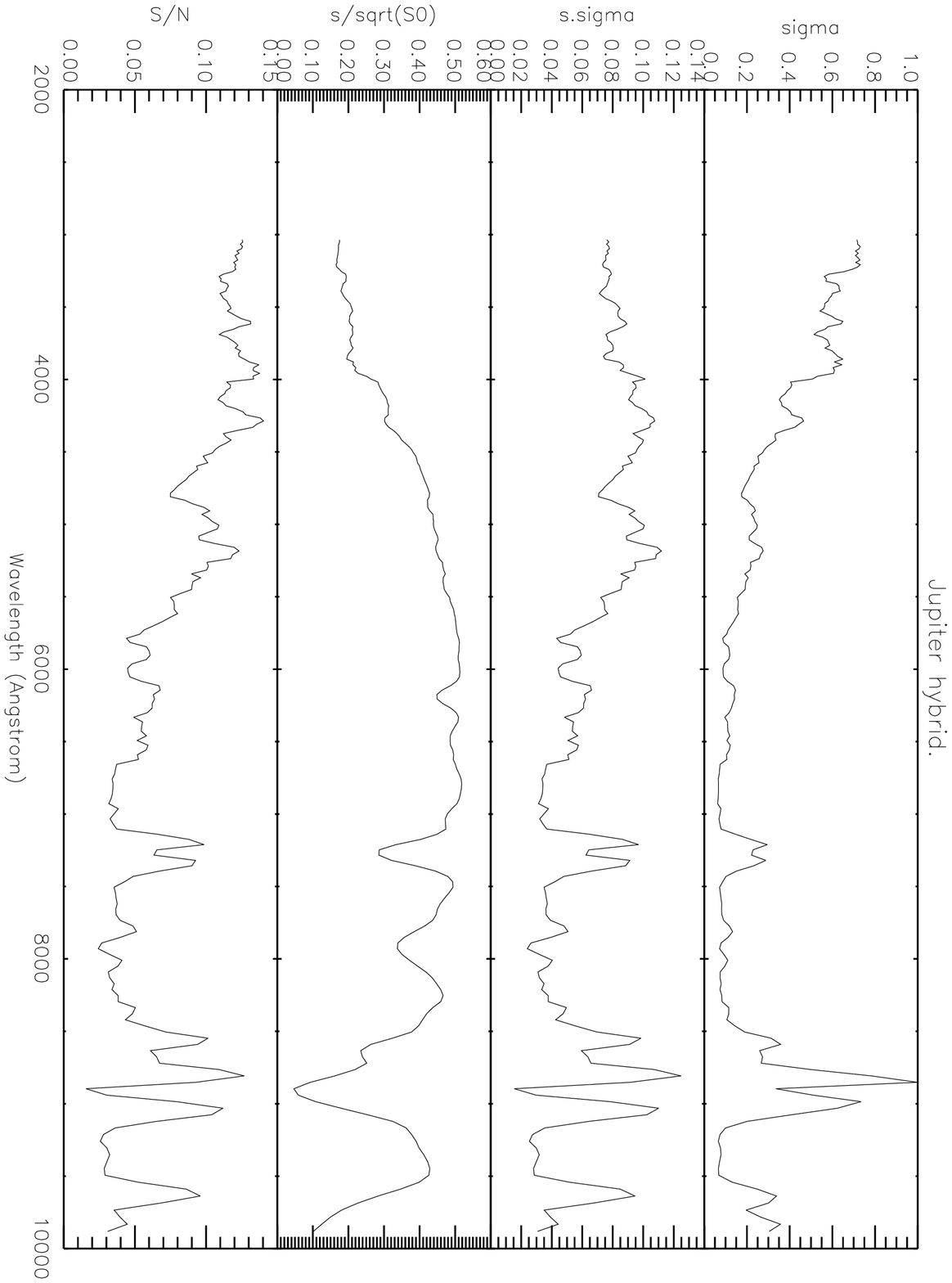}
\end{figure}

\clearpage

\begin{figure}
\plotone{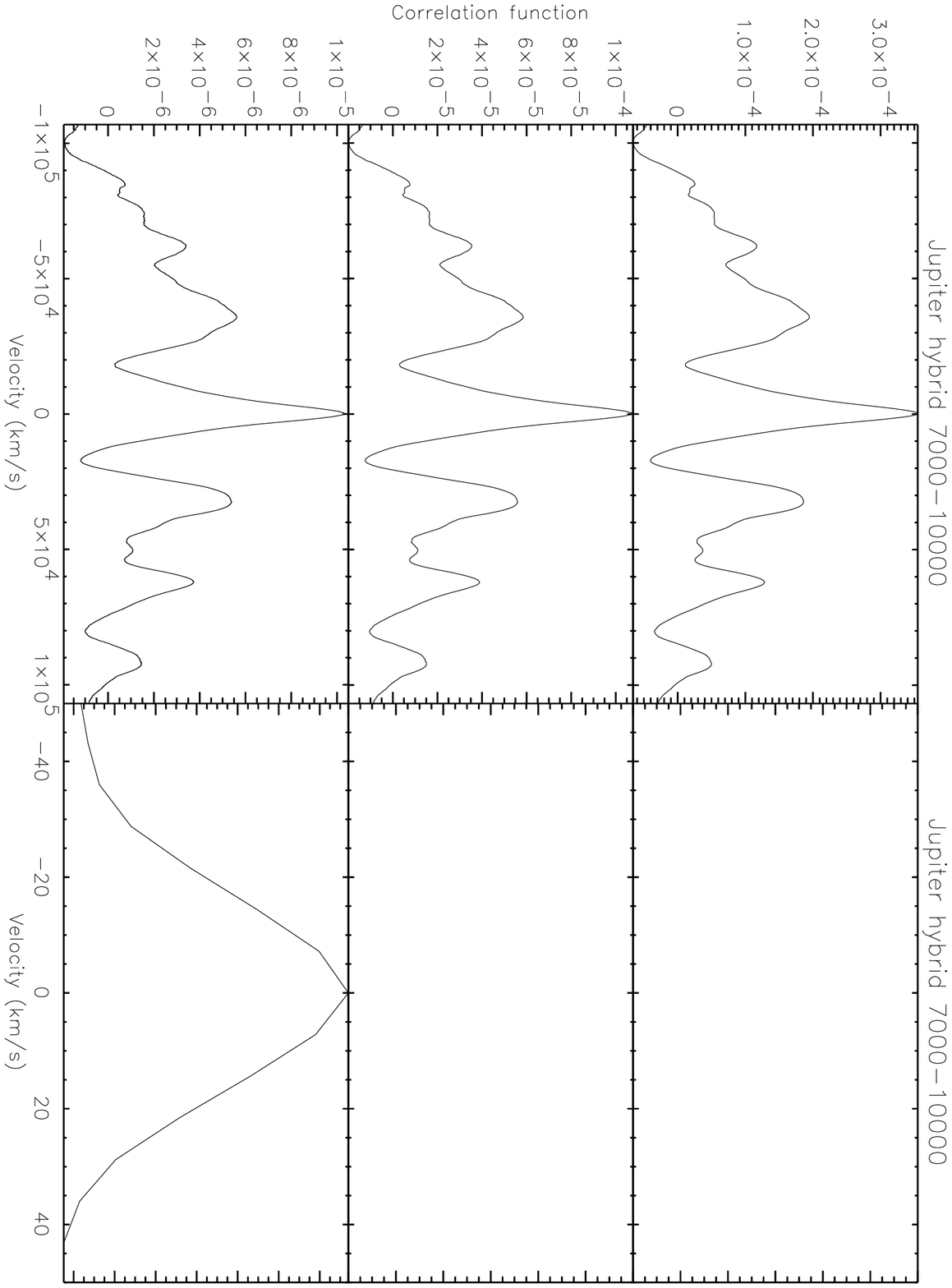}
\end{figure}

\clearpage

\begin{figure}
\plotone{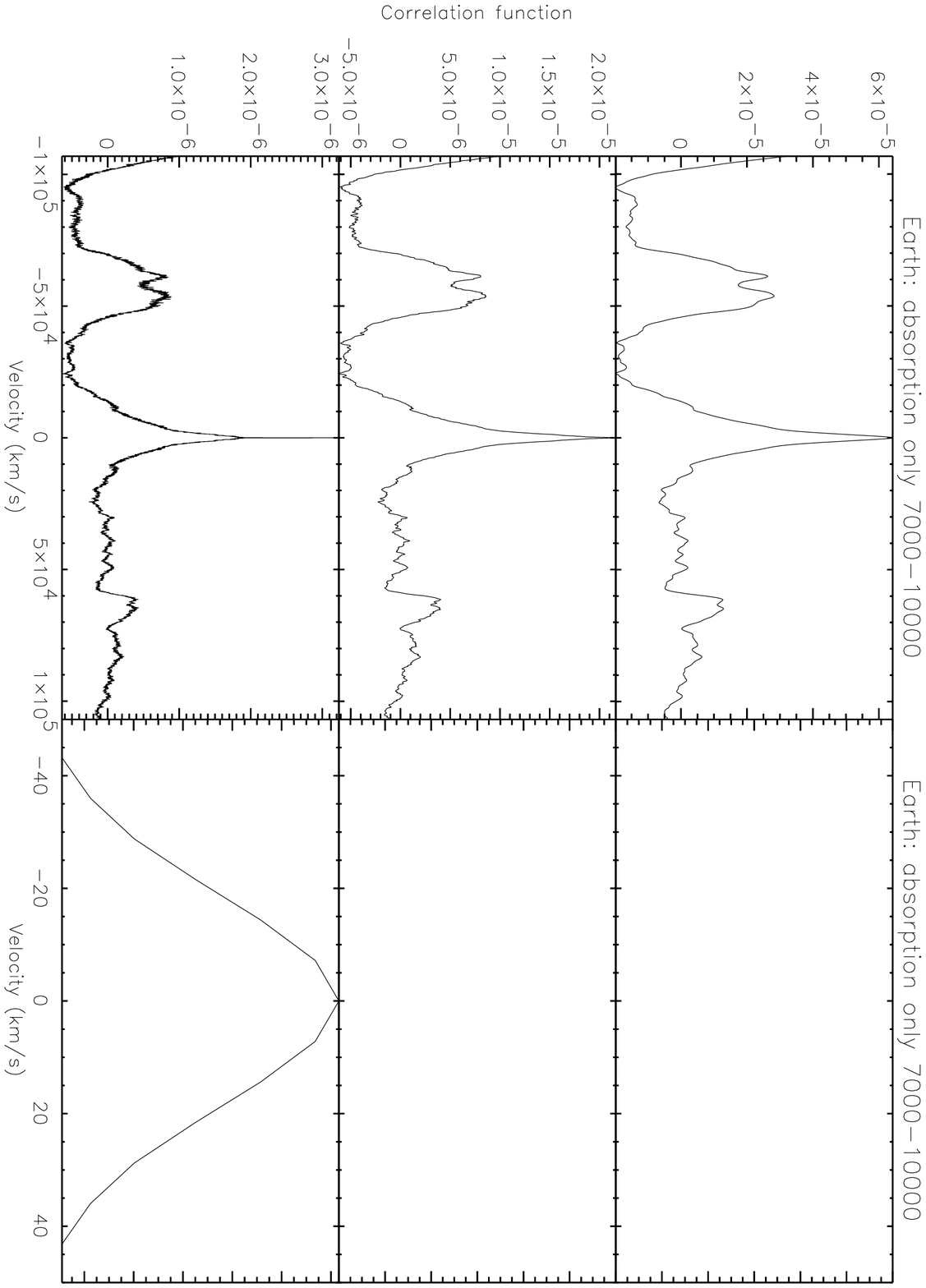}
\end{figure}

\clearpage

\begin{figure}
\plotone{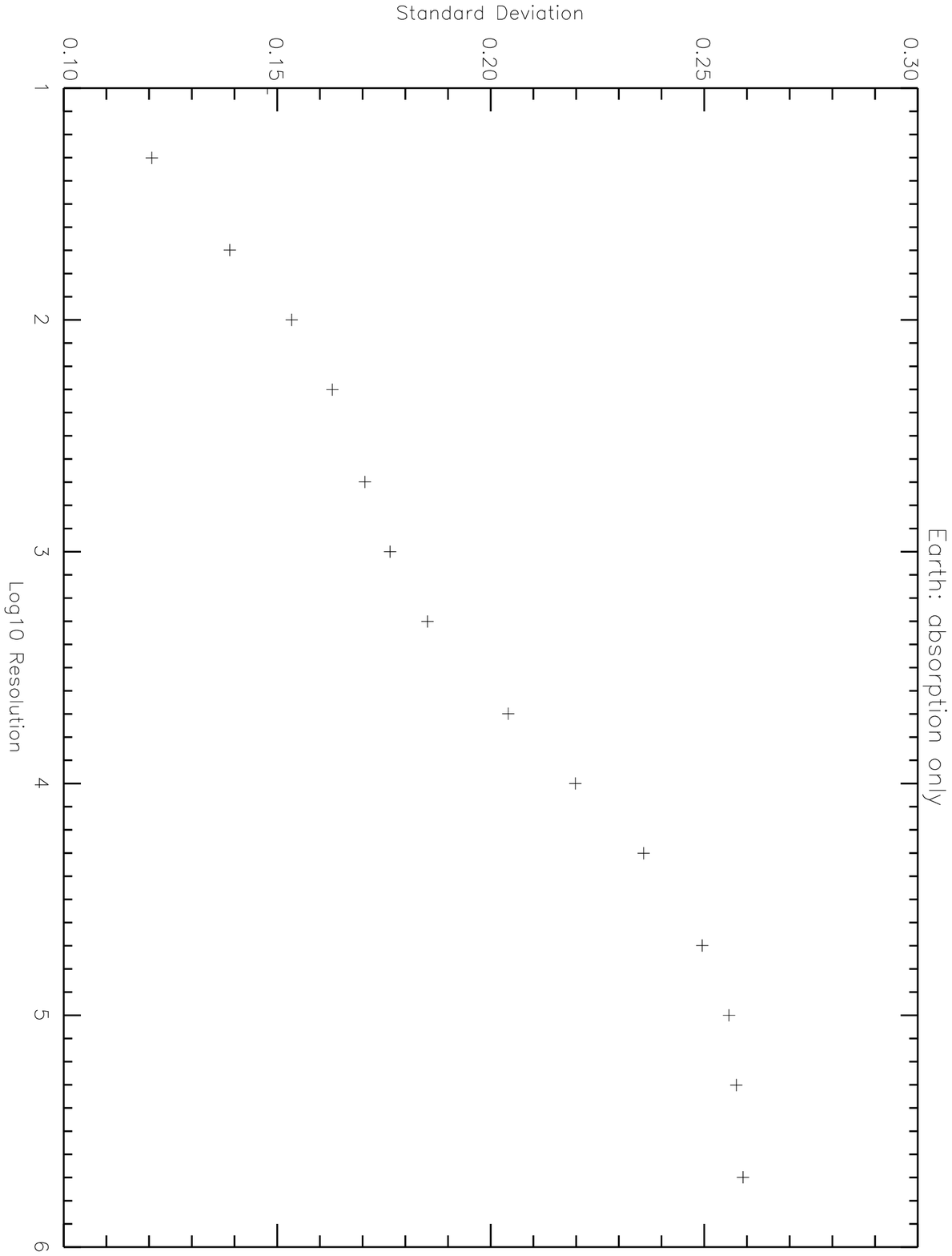}
\end{figure}

\clearpage

\begin{figure}
\plotone{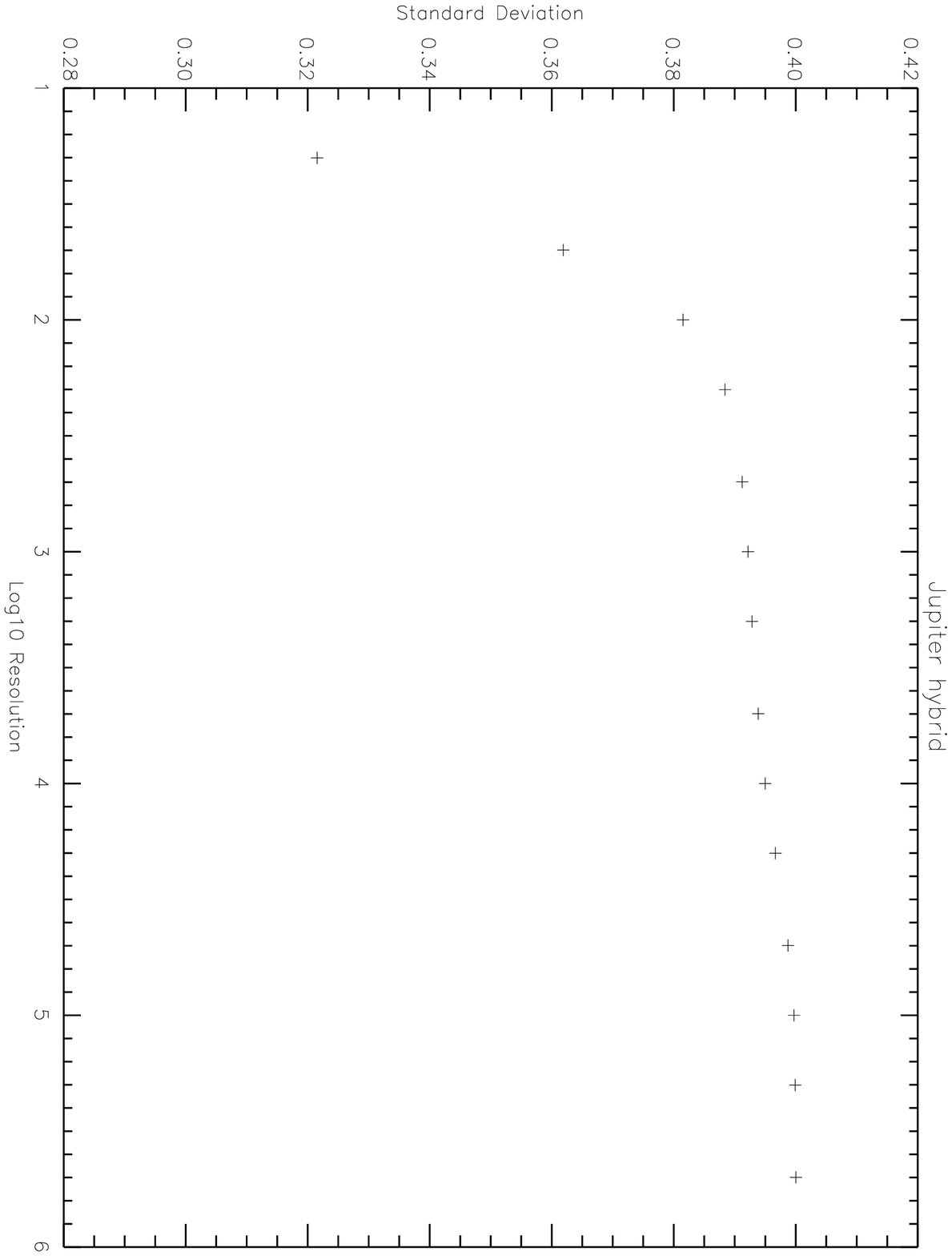}
\end{figure}

\clearpage

\begin{figure}
\plotone{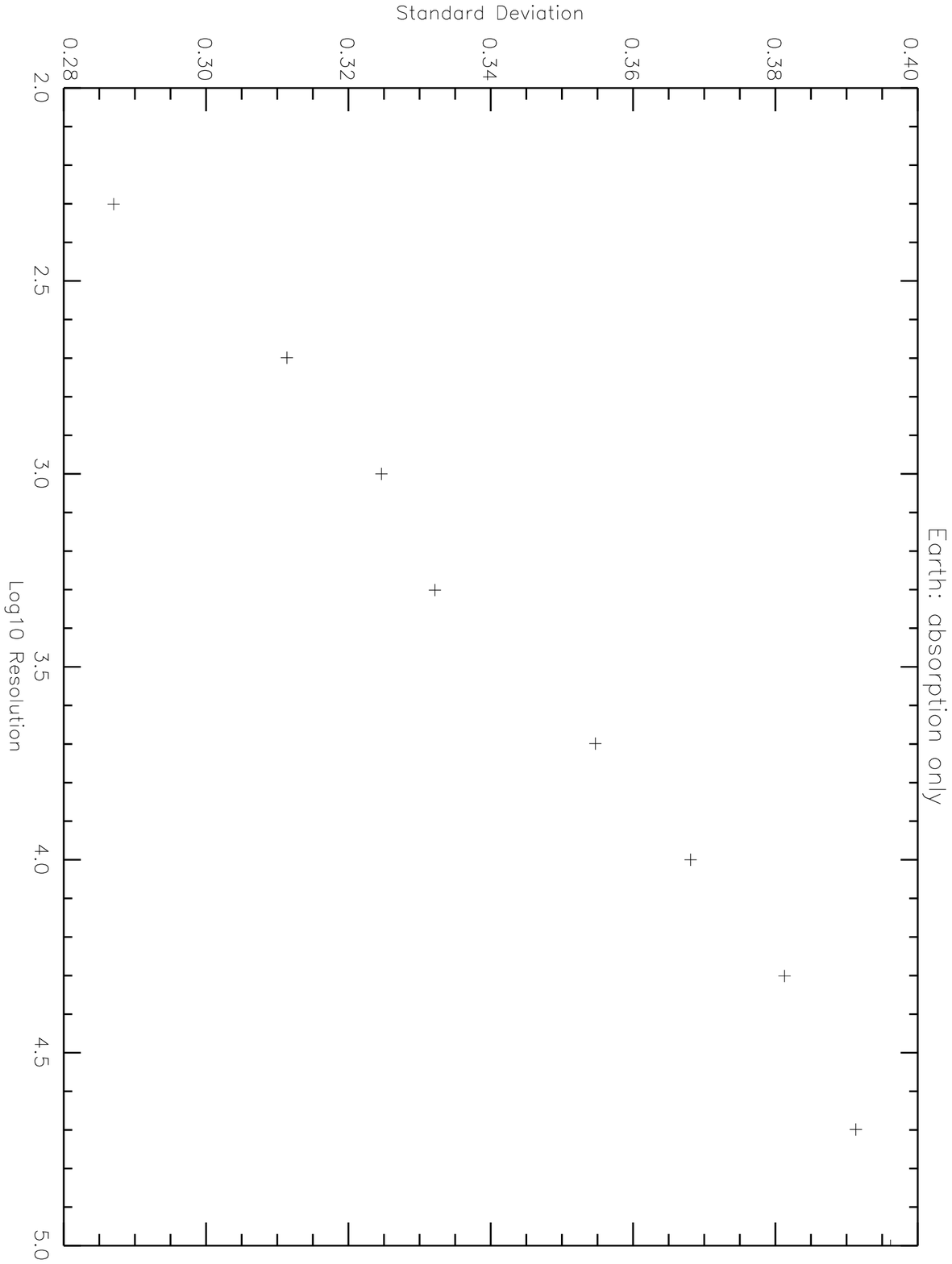}
\end{figure}

\clearpage

\begin{figure}
\plotone{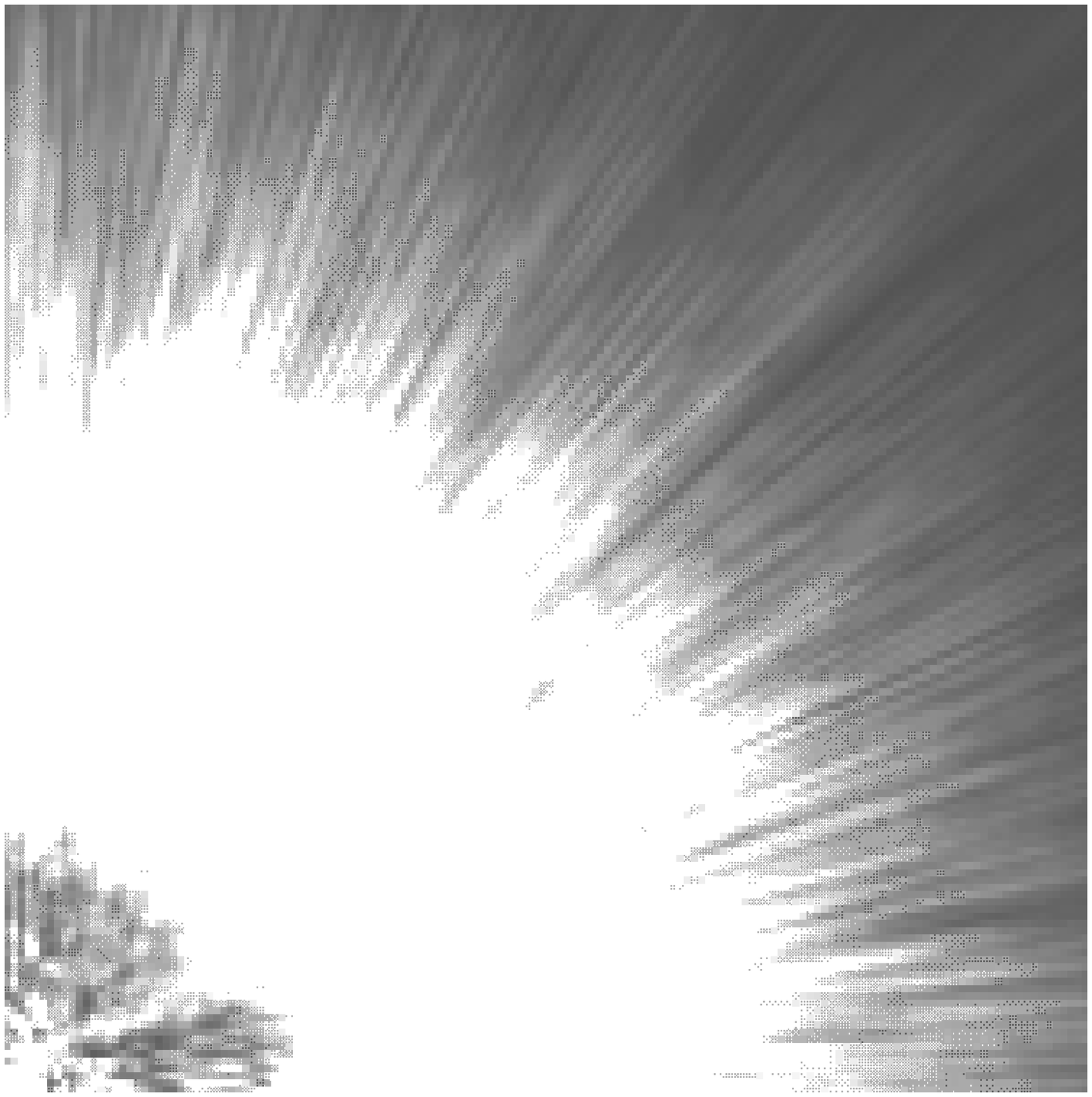}
\end{figure}

\clearpage

\begin{figure}
\plotone{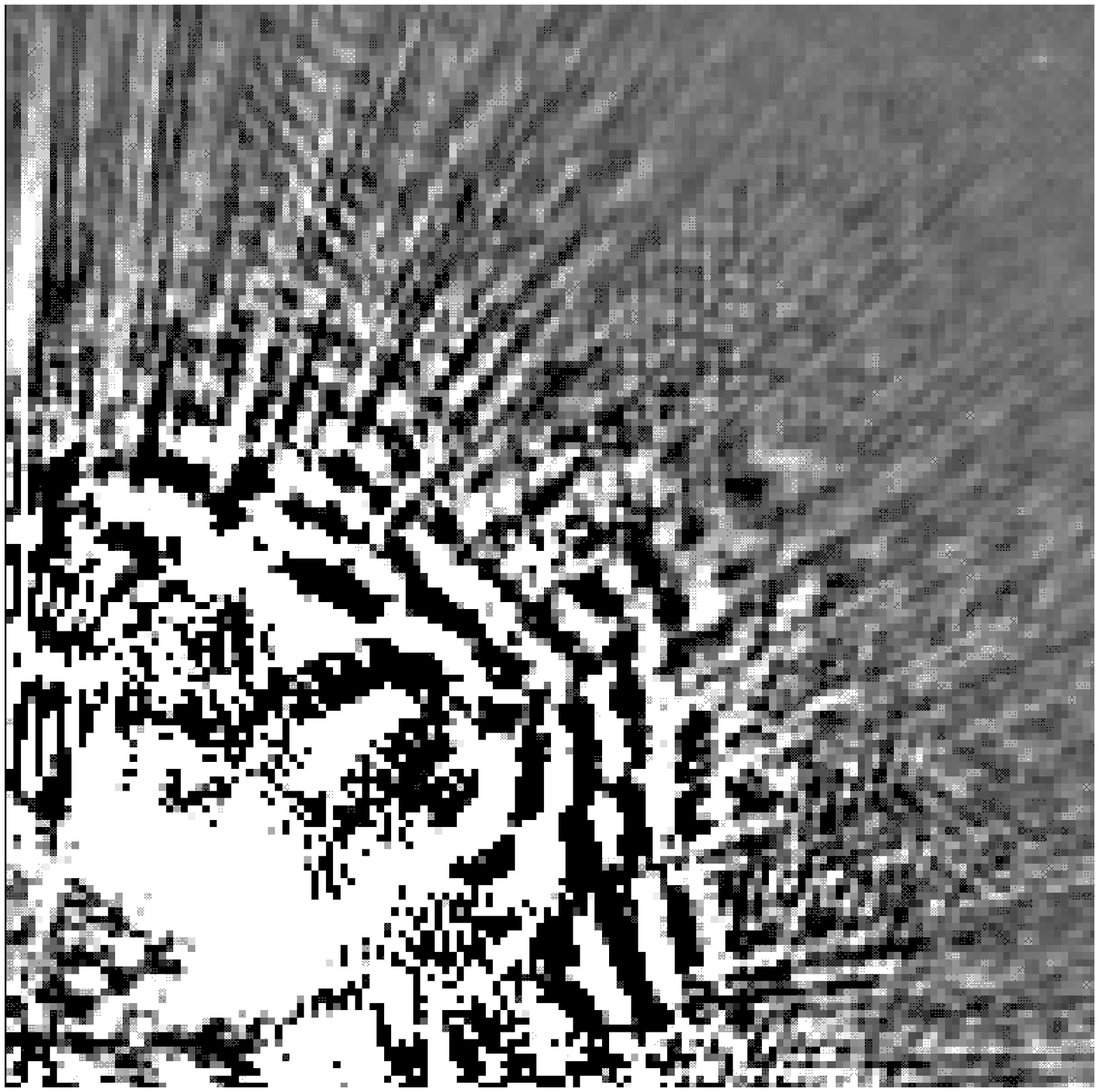}
\end{figure}

\clearpage

\begin{figure}
\plotone{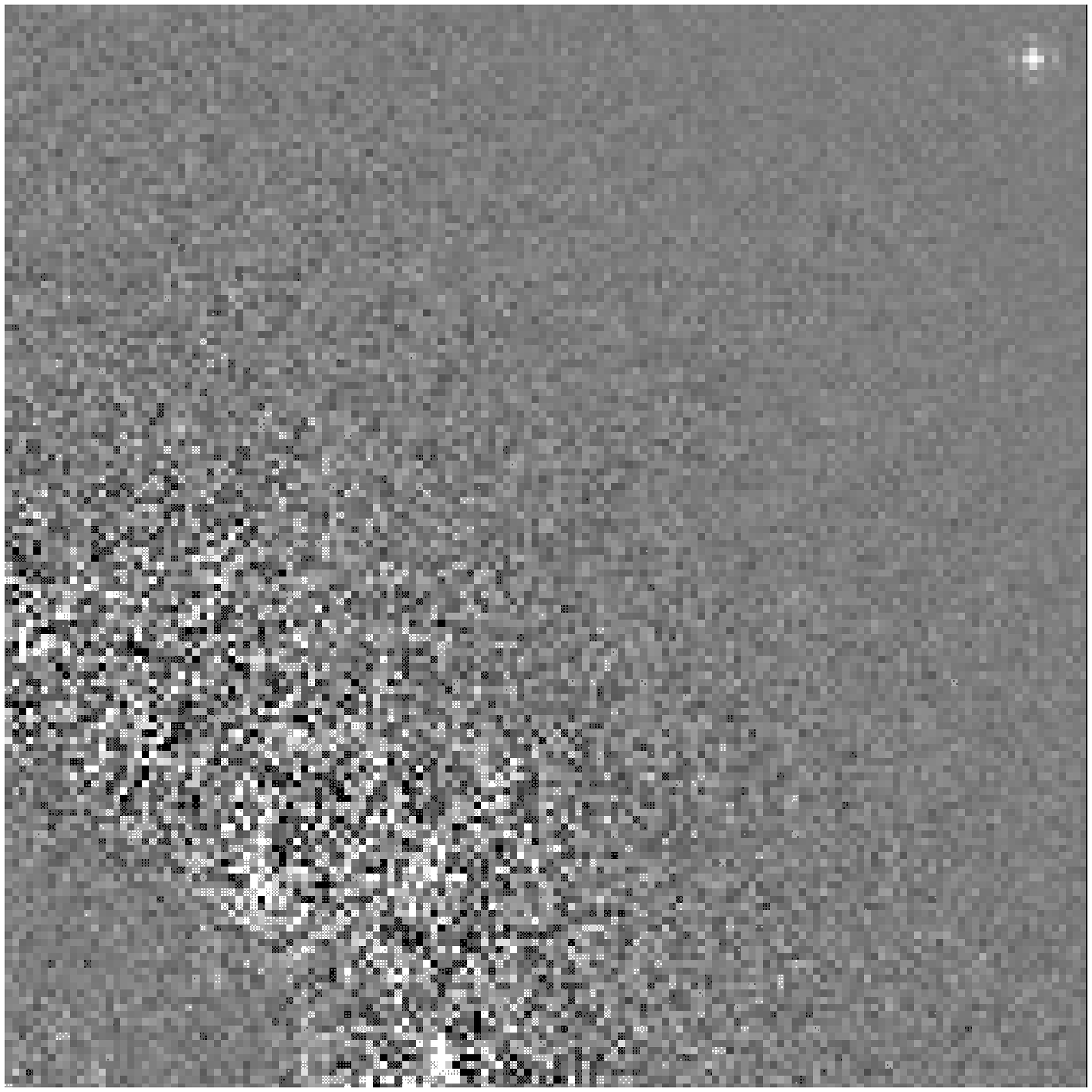}
\end{figure}

\clearpage

\begin{figure}
\plotone{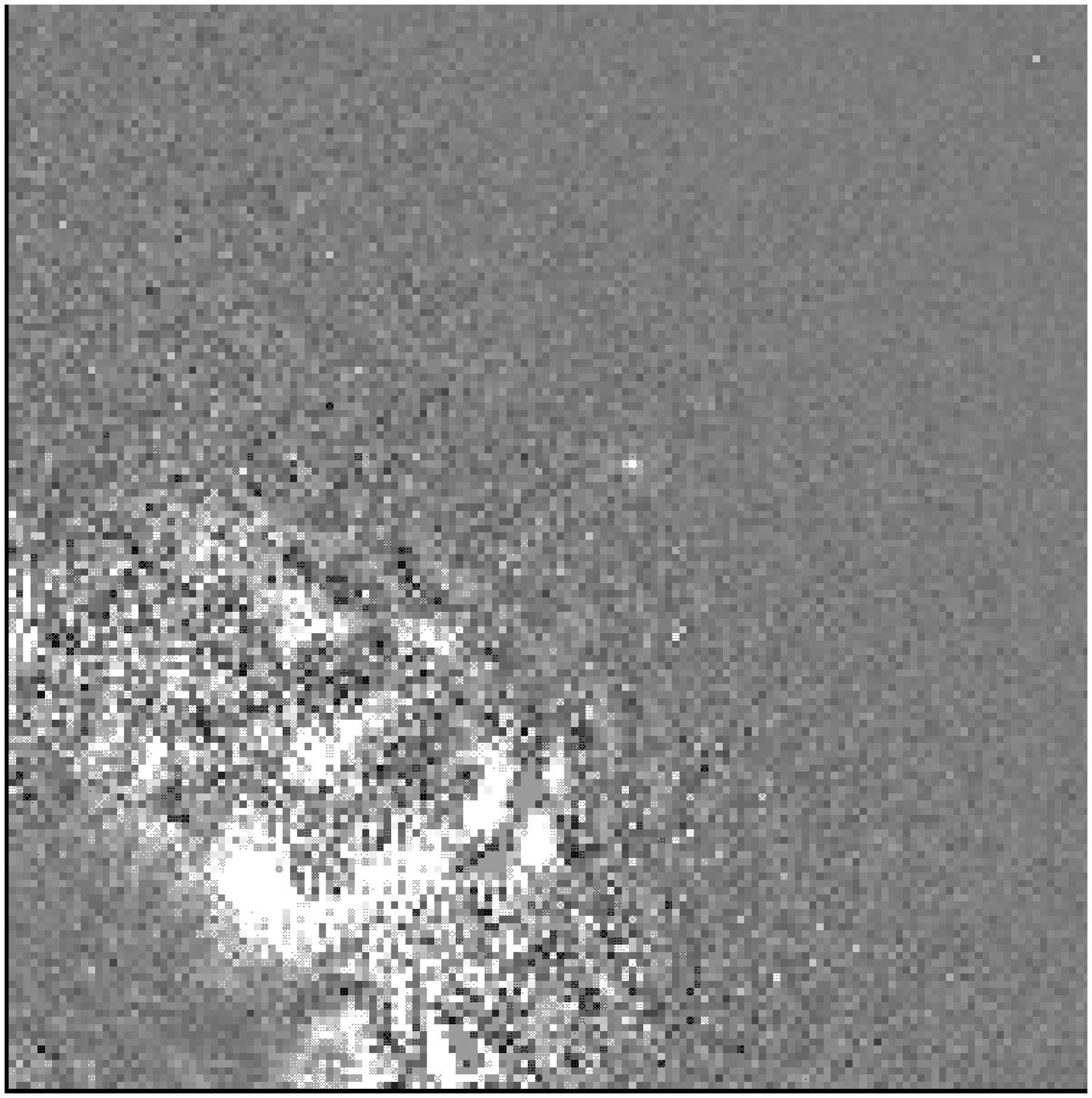}
\end{figure}

\clearpage

\begin{figure}
\plotone{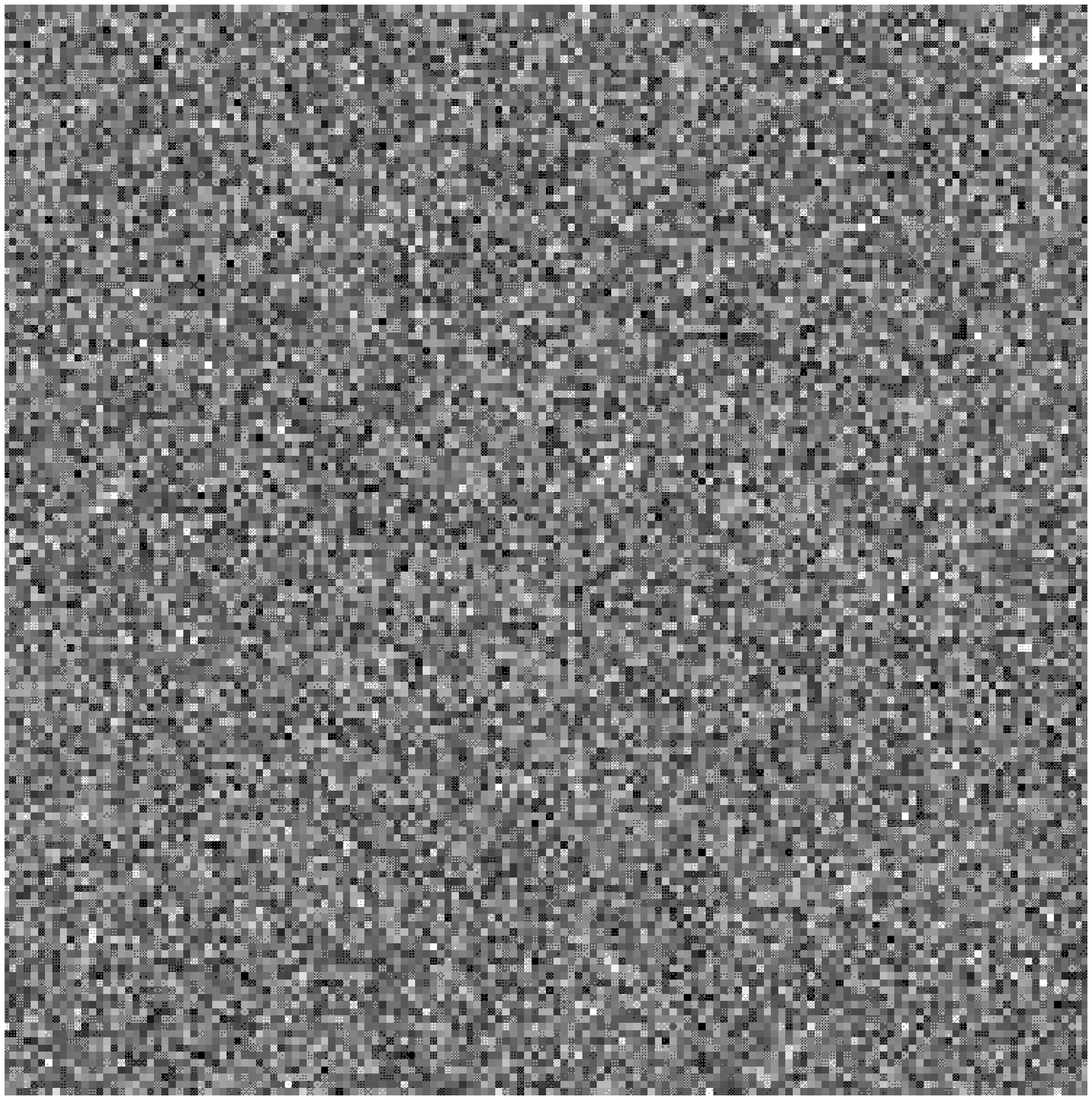}
\end{figure}

\clearpage

\begin{figure}
\plotone{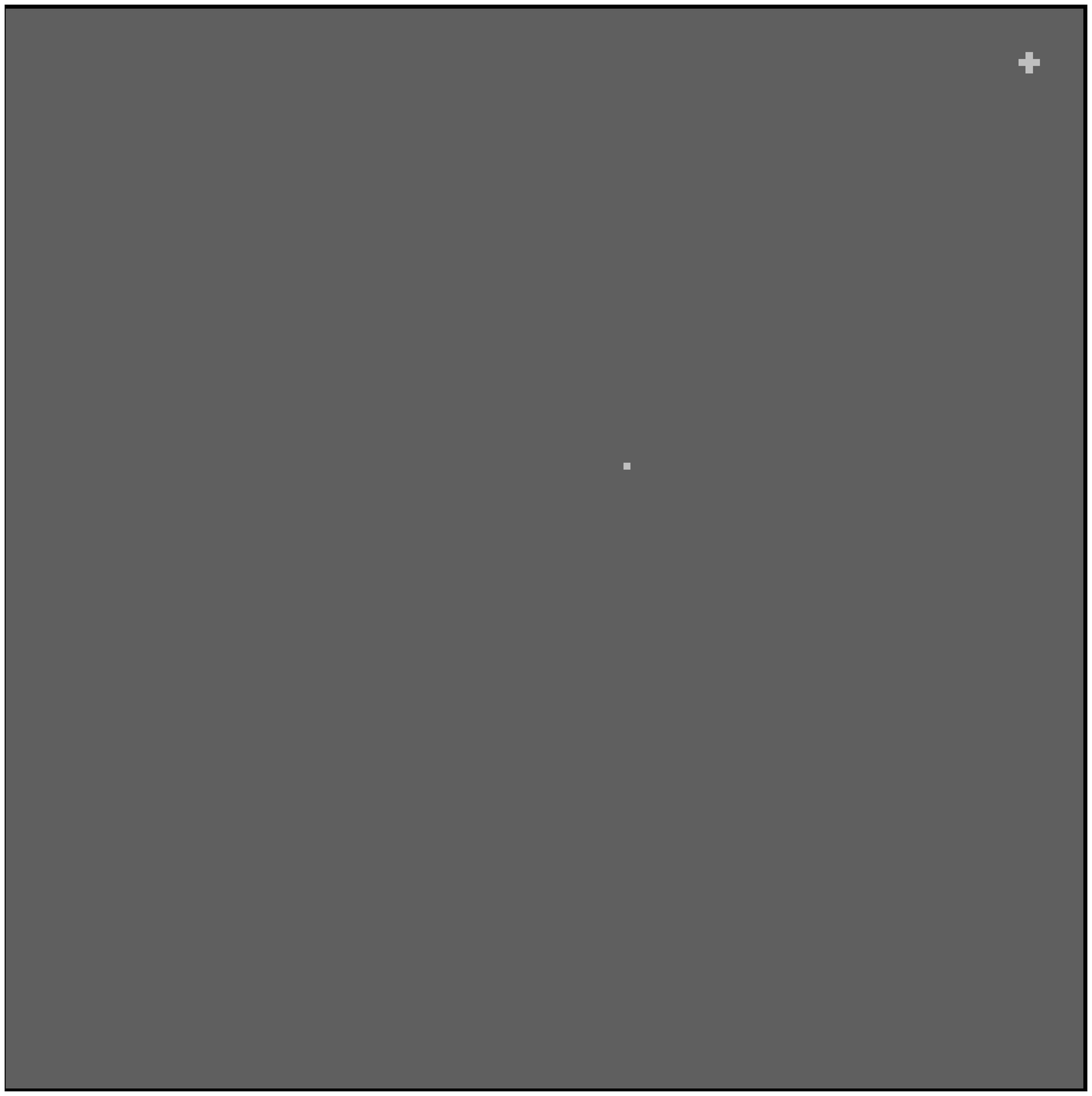}
\end{figure}

\clearpage

\begin{figure}
\plotone{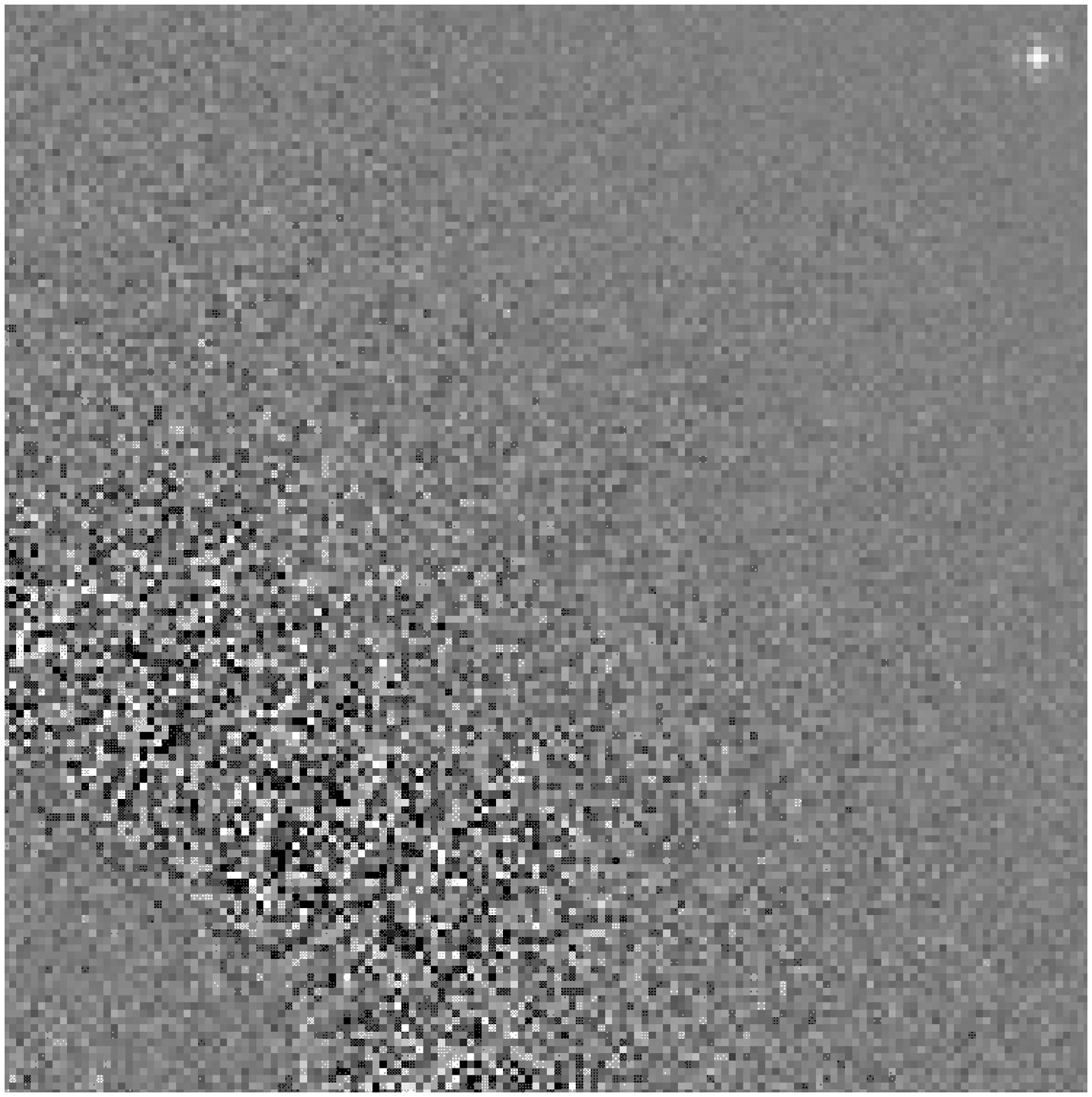}
\end{figure}

\clearpage

\begin{figure}
\plotone{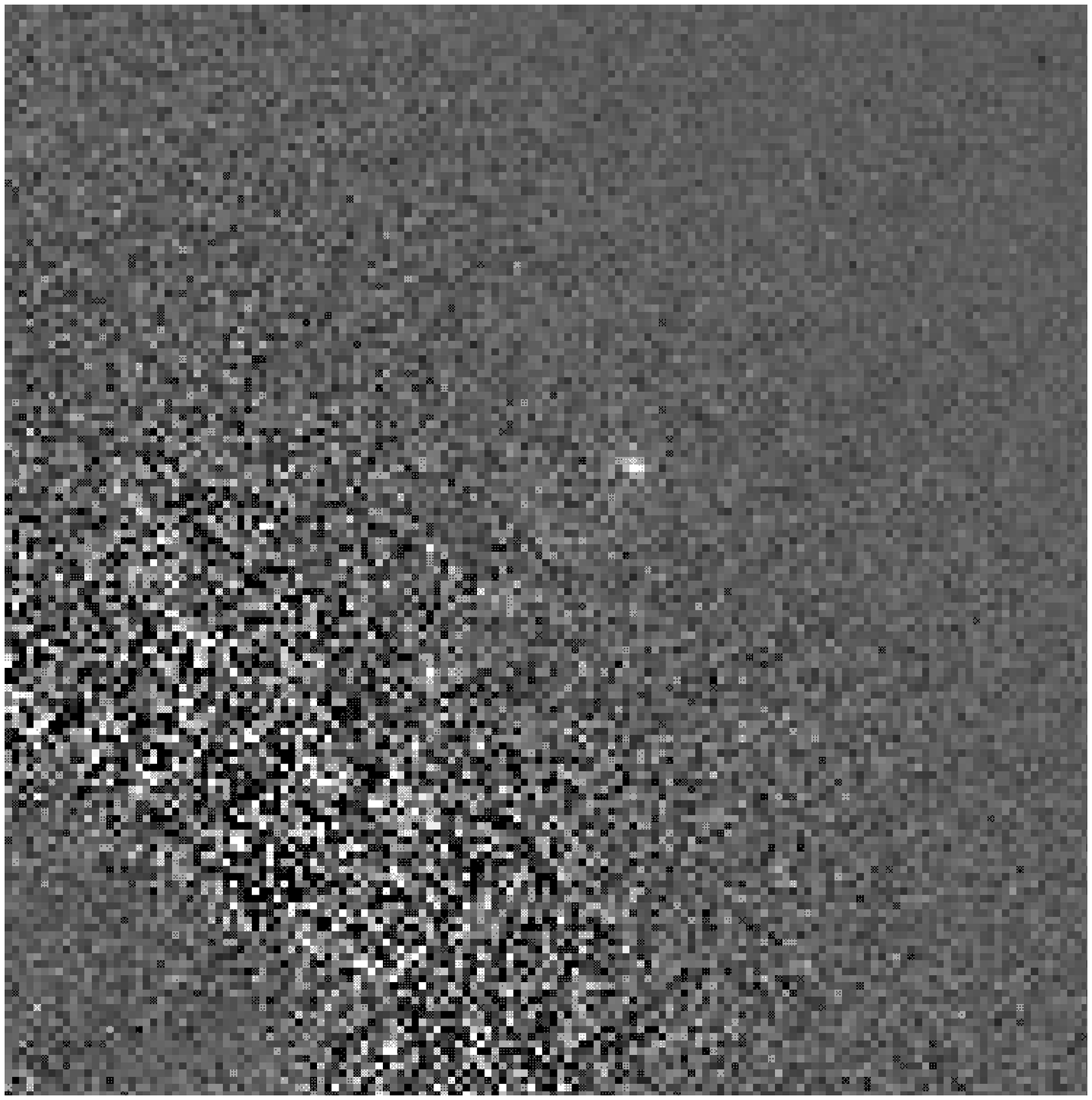}
\end{figure}

\begin{figure}
\plotone{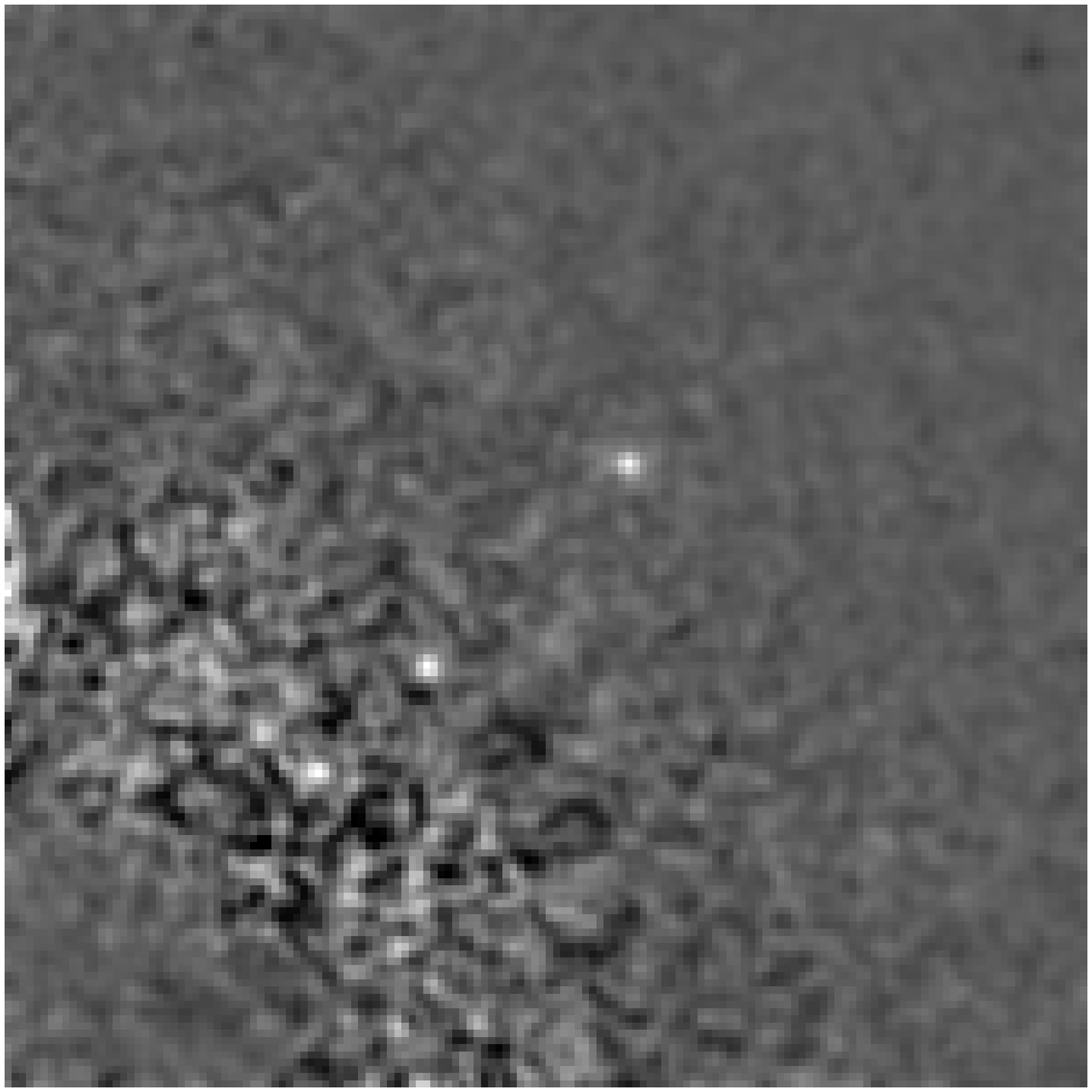}
\end{figure}

\begin{figure}
\plotone{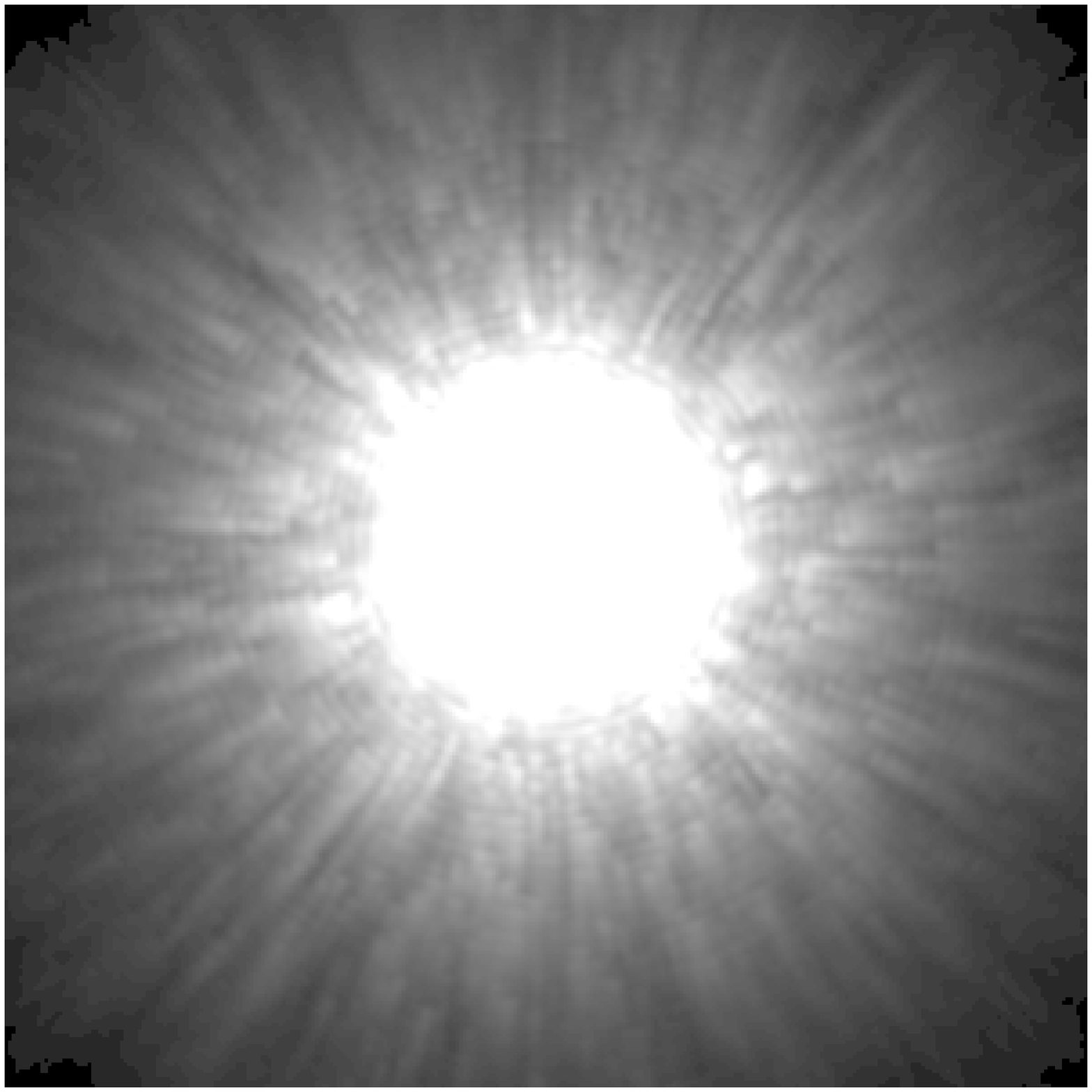}
\end{figure}

\begin{figure}
\epsscale{0.7}
\plotone{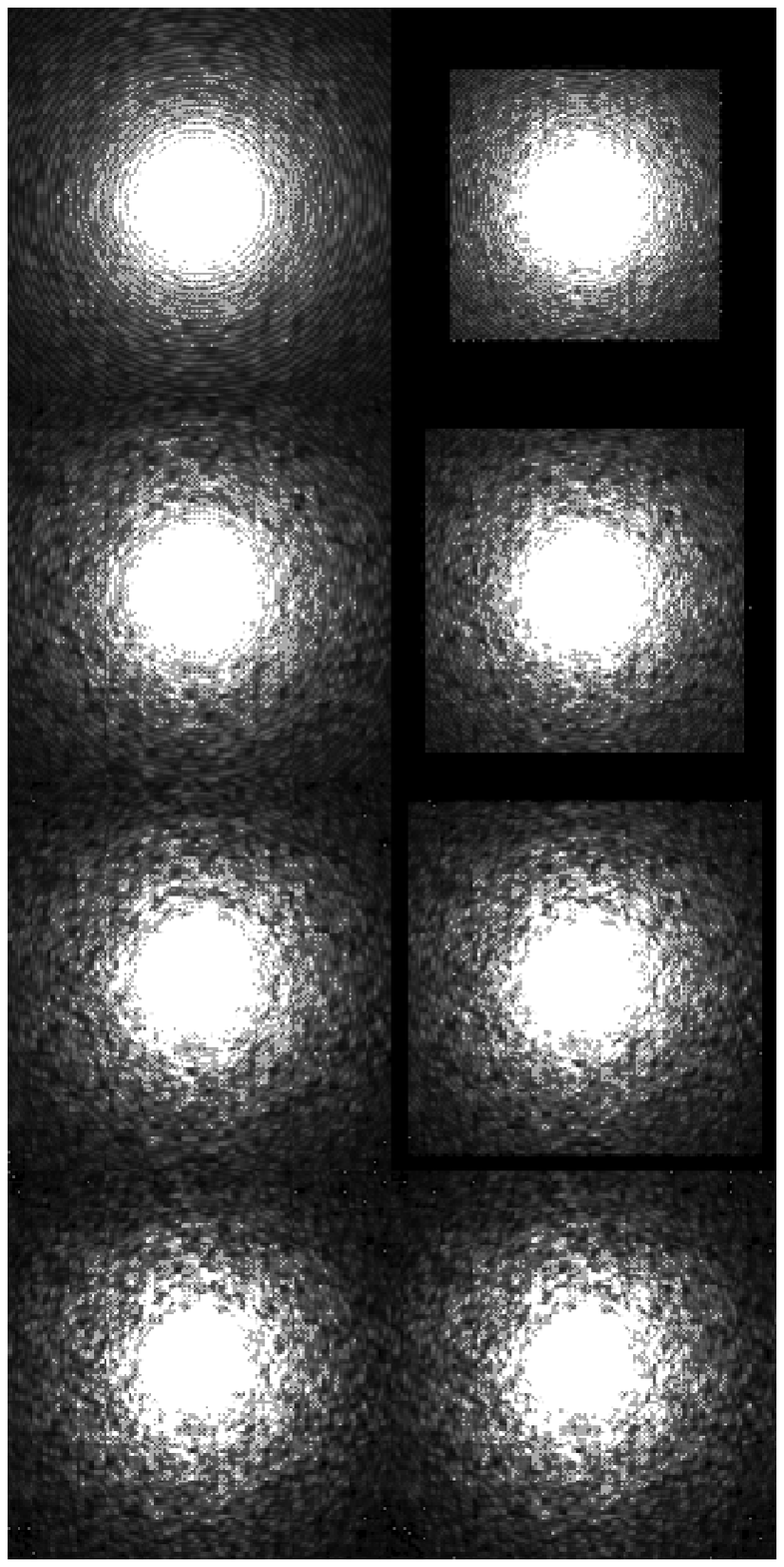}
\epsscale{1.0}
\end{figure}

\begin{figure}
\plotone{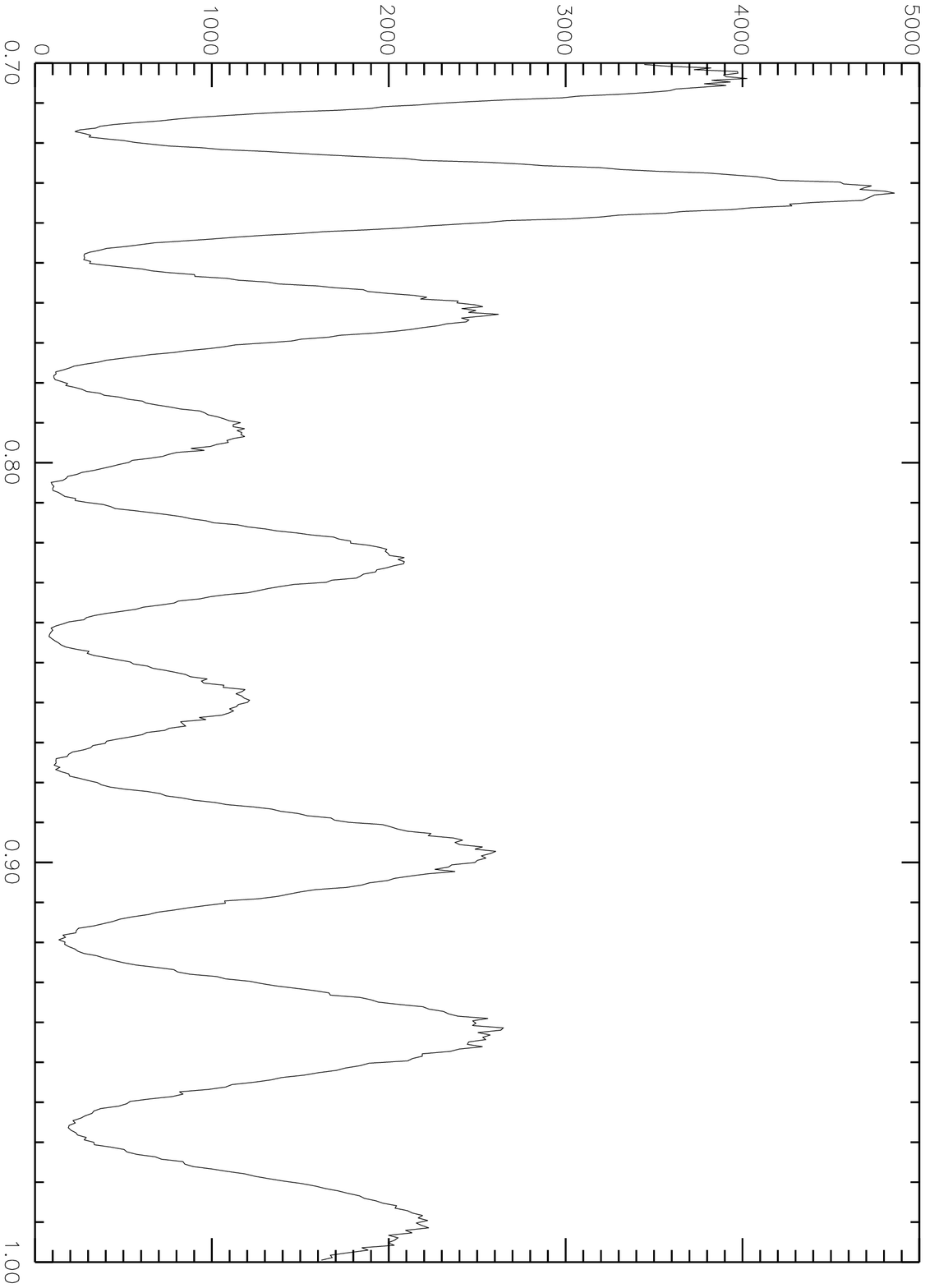}
\end{figure}

\begin{figure}
\plotone{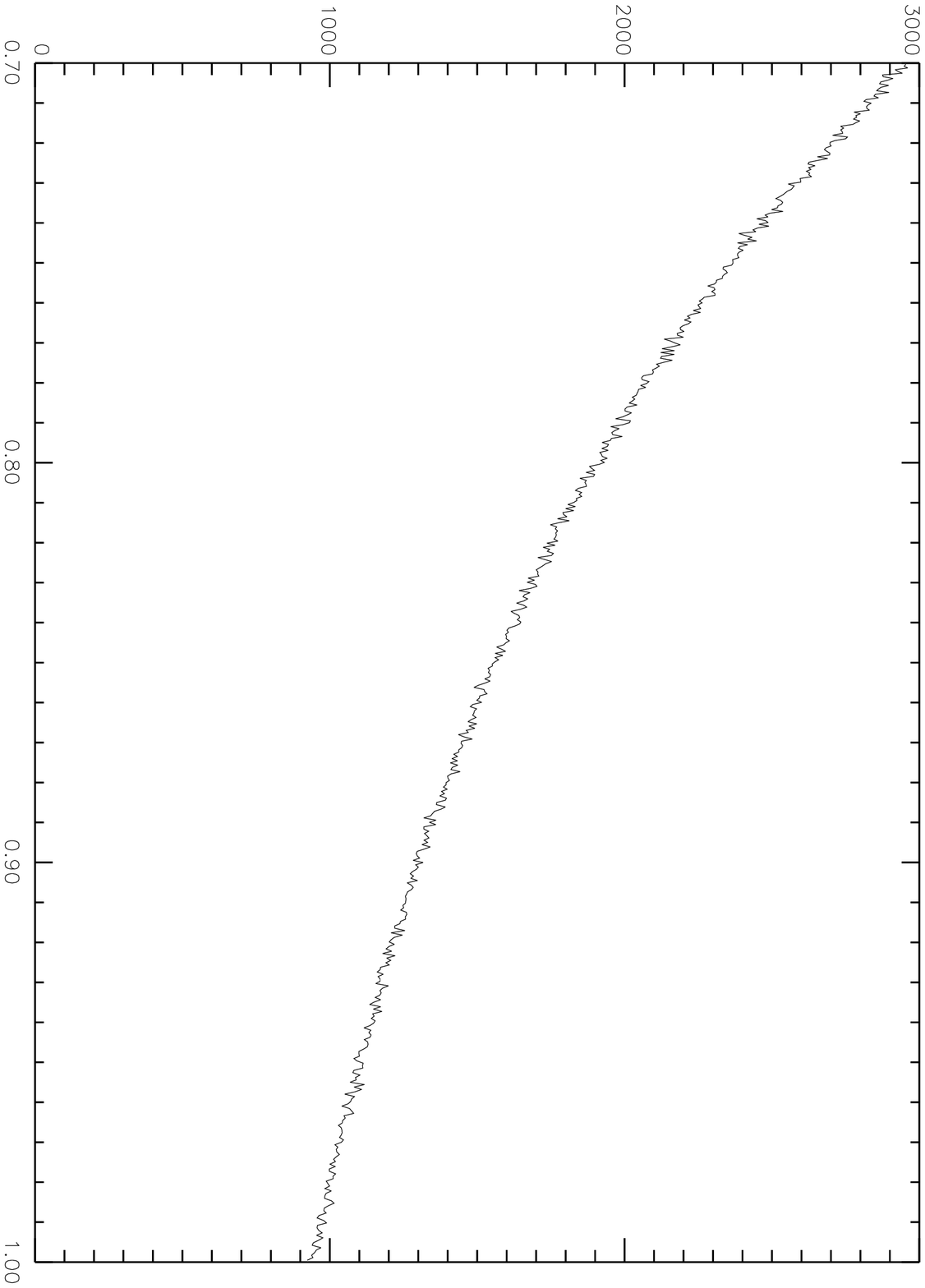}
\end{figure}

\begin{figure}
\plotone{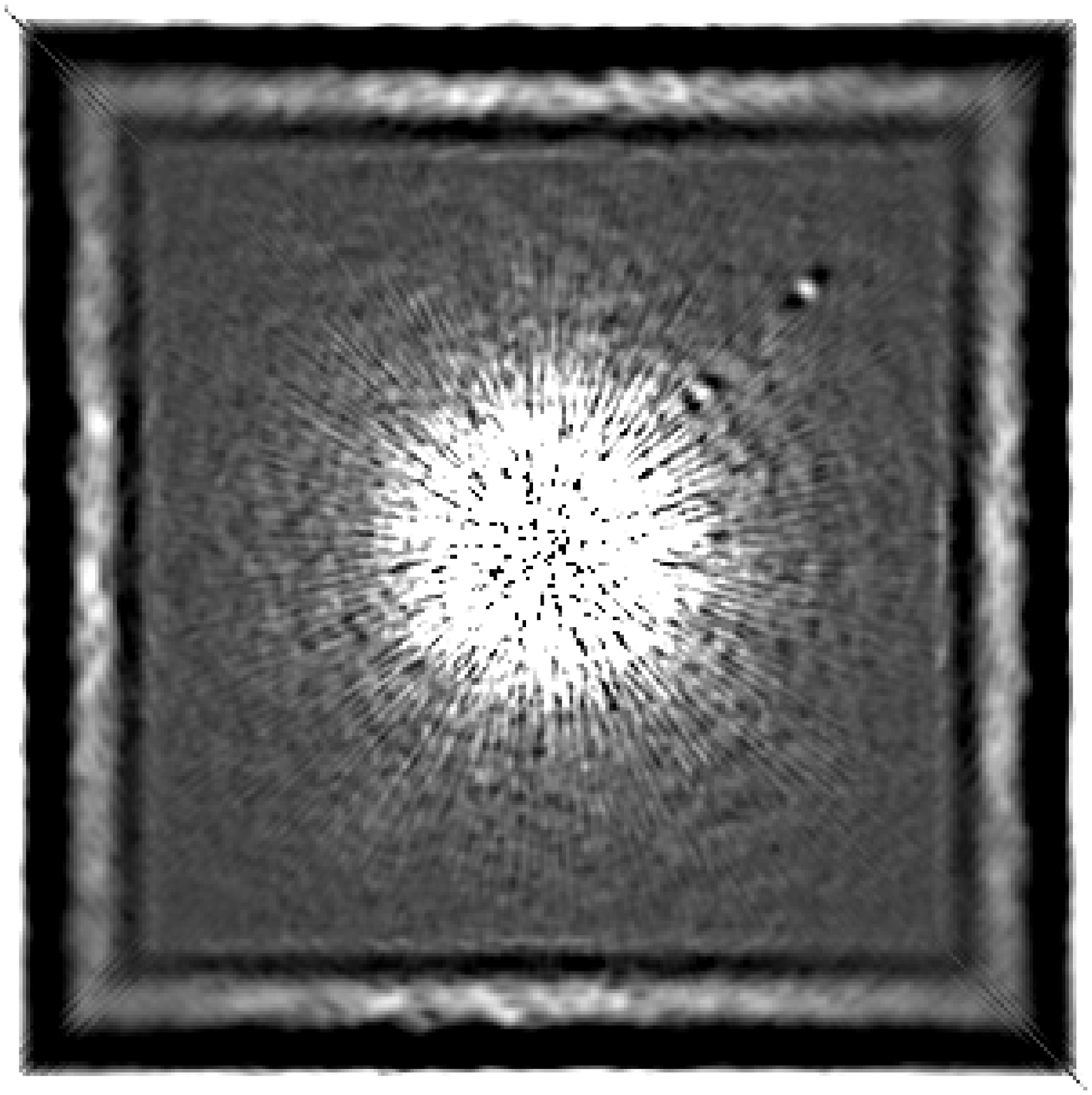}
\end{figure}

\begin{figure}
\plotone{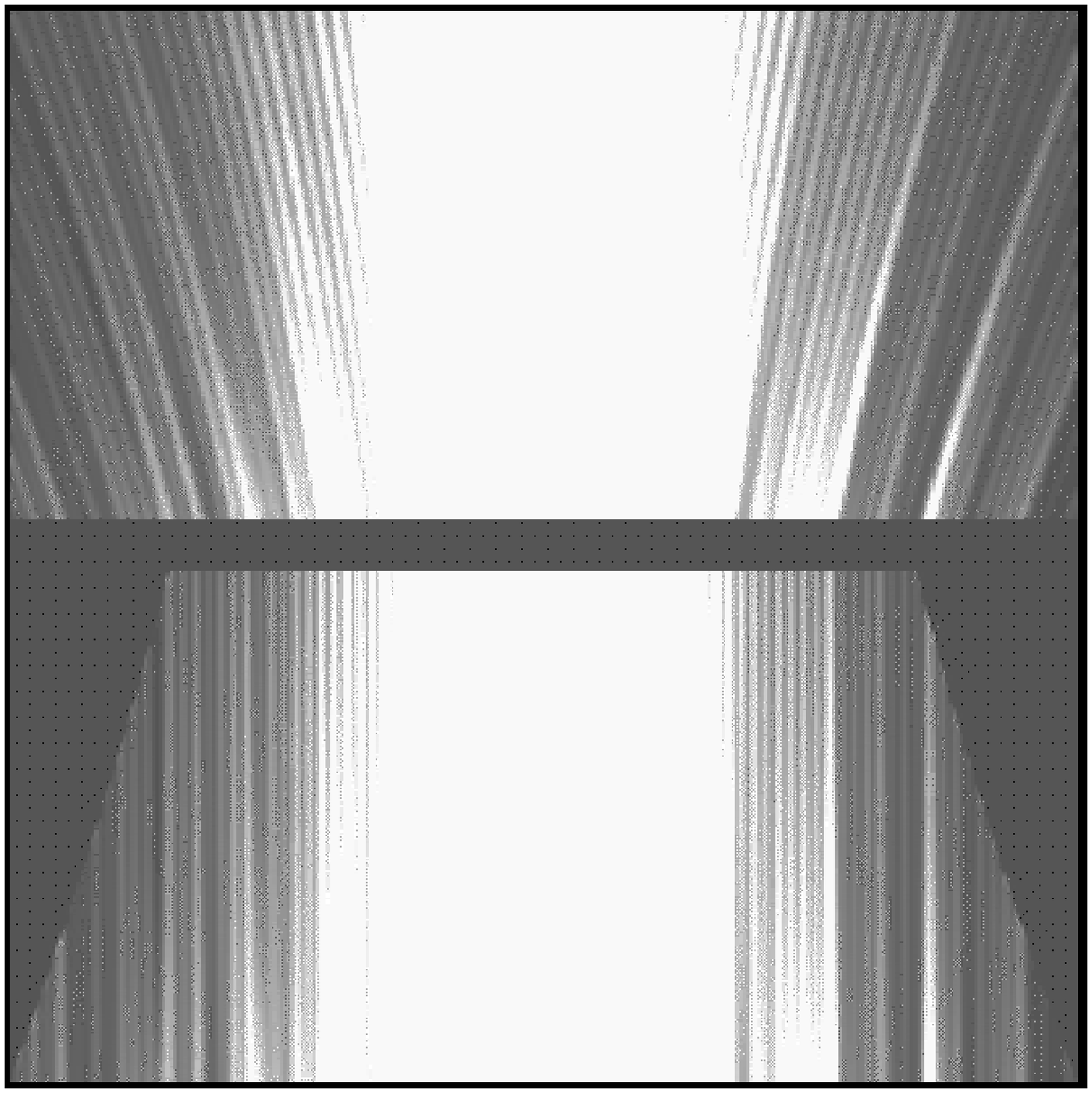}
\end{figure}



\figcaption{Solar spectrum, continuum normalized, showing the growth of structure
as spectral resolution increases.
Although the spectra in this plot and the following are plotted against wavelength,
the actual underlying sampling is logarithmic. The resolution is as indicated in the panels.
\label{fig1}}

\figcaption{Hybrid Jovian template constructed from high resolution
Solar model multiplied by moderate resolution Jovian albedo spectrum,
Karkoschka (1994).}

\figcaption{Terrestrial template obtained by multiplying the high resolution
Solar model by the square of the atmospheric transmission functions
of the empirical Solar atlases described in the text.}

\figcaption{Detail from the previous figure, showing the oxygen A-band
absorption feature, and the growth of fine structure as the spectral
resolution is increased.}

\figcaption{Similar terrestrial template to Fig.~3, but now with
OH emission lines in the region $1 \mu m$ --- $2 \mu m$ added
as described in the text to
assess their observability at high resolution.}

\figcaption{For a spectral window 2\%\ of the wavelength in question,
we plot the cross-correlation
detectability parameters (Equ.~5) in various combinations
as a function of wavelength for the terrestrial
template. The top panel plots the standard deviation of the unit-mean template,
$\sigma$, the second panel is the product $s_p \sigma$ with $s_p$ from the template.
The
third panel is $s_p/\sqrt{S_0}$ with $S_0$ from the stellar template,
and the bottom panel the final
$S/N = s_p\sigma/\sqrt{S_0}$. }

\figcaption{For a spectral window 2\%\ of the wavelength in question,
we plot the cross-correlation
detectability parameters as a function of wavelength for the Jovian
template with panels as in the caption to Fig.~6.}

\figcaption{Noiseless autocorrelation function for the Jovian template in the
wavelength region 7000\AA\ --- 10,000\AA\ as a function of spectral resolution.
The top, center and bottom panels are for spectral resolution 300, 1000 and
10,000 respectively.
There is a minor increase in the sharpness of the autocorrelation function as the resolution
is increased, but most of the power in the spectrum comes from the broad,
diffuse bands (which may be an artifact of the template construction).
The right hand column expands the velocity scale, intended to show the high resolution
case autocorrelation peak.}

\figcaption{Noiseless autocorrelation function for the terrestrial template
in the
wavelength region 7000\AA\ --- 10,000\AA\ as a function of spectral resolution.
The top, center and bottom panels are for spectral resolution 300, 1000 and
10,000 respectively.
Because of the dominance of fine structure in the terrestrial spectrum, the
autocorrelation becomes very finely peaked in the high resolution case.
As in Fig.~8, the right hand column expands the velocity scale about the
autocorrelation peak to illustrate the accuracy of velocity measurement
that might be achieved.
}

\figcaption{Dispersion of the unit-mean template as a function of spectral
resolution for the terrestrial template from 7000\AA\ to 10,000\AA .
Note the continuous rise in $\sigma$ due to the presence of fine structure
in the spectrum.}

\figcaption{Dispersion of the unit-mean template as a function of spectral
resolution for the Jovian template.
The standard deviation ceases to rise at resolutions above $\approx 300$
as the majority of power in this template lies in the diffuse molecular
band features. This may be an artifact of the template.}

\figcaption{Dispersion of the unit-mean template as a function of spectral resolution for the terrestrial template from 7580\AA\ to 7800\AA ,
corresponding to a spectral region including the oxygen A-band complex.
The high level of complexity in this feature causes the standard
deviation to rise continuously with resolution.}

\figcaption{Simulation of realistic coronagraphic image obtained using the
ACS aberrated beam coronagraph showing typical PSF wings and is intended
to provide a generic representation of PSF complexity that might be
encountered.}

\figcaption{Simulation of a ``roll deconvolution'' in which a pair of
coronagraphic images obtained at different roll angles have been subtracted
from one another. Slight miscenterings and focus changes (breathing) were introduced between
the two. There is a Jovian planet visible in the top right corner.}

\figcaption{Correlation function, $c_0$, for the roll-deconvolved datacube
using the hybrid Jovian template. The Jovian planet is clearly visible in the top right corner.}

\figcaption{Correlation function for the roll-deconvolved datacube
using the terrestrial template. The ``bright Earth'' is clearly visible.}

\figcaption{Analysis of variance image showing an image of reduced-$\chi^2$
with a peak at the location of the Jovian planet, and a second at the inner
position of the ``bright Earth''.}

\figcaption{Image of the probability associated with reduced-$\chi^2$ thresholded
at $5-\sigma$ showing a statistical signature of both planets.}

\figcaption{Gram-Schmidt correlation function image using the Jovian template
applied to the original (not roll-deconvolved) coronagraphic datacube. The Jovian
planet is clearly visible, and the remainder of the image shows only noise
expected from photon statistics.}

\figcaption{Gram-Schmidt correlation function image using the terrestrial
template
applied to the original (not roll-deconvolved) coronagraphic datacube. The 
outer bright-Earth
planet is clearly visible.}

\figcaption{Lightly smoothed version of the previous figure, showing all
three inner planets. Note that the outer Jovian planet does not appear.}

\figcaption{Simulation of space-based coronagraphic observation of Solar type
star at a distance of 2~pc over the wavelength range 7000\AA\ to 1$\mu$m, with parameters as
given in the text. The figure shows the spectrally integrated PSF, representative
of a direct image taken with an I-band filter.}

\figcaption{Sequences showing cuts through the datacube at wavelengths
0.70~$\mu$m, 
0.76~$\mu$m, 
0.84~$\mu$m and
1.0~$\mu$m, bottom to top,
The left four panels show the speckles and ripples expanding with wavelength.
The right panels illustrates the process of adjusting image scale as a function
of wavelength, and PSF features can now be seen to remain fixed in place.}

\figcaption{Spectral plot for a pixel near the Jovian planet in the original datacube
showing strong modulation as Airy rings and speckles move across the pixel.}

\figcaption{Spectral plot for the same pixel in the scale-adjusted datacube
showing the removal of rapid modulation and only low frequency terms which are
easily modelled.}

\figcaption{Spectrally collapsed image of the reconstructed, subtracted datacube
showing a $20\sigma$ detection of a Jovian planet whose peak is only
1\%\ of the PSF wing background, using a 2-m telescope.
The equivalent observation with an 8-m TPF would 
correspond to a terrestrial scale planet.}

\figcaption{Cuts through the original datacube with one spatial and one spectral dimension.
In the original cube, the rings diverge outwards with wavelength, while in the spatially resampled cube,
they form straight lines running vertically in the plot and
showing only low frequency structures.}

\clearpage





\clearpage

\begin{deluxetable}{crrrr}
\tabletypesize{\scriptsize}
\tablecaption{Photon counts for extrasolar planets.}
\tablewidth{0pt}
\tablehead{
\colhead{Telescope diameter} &
\colhead{Distance (pc)}   &
\colhead{Jupiter}   &
\colhead{Earth} &
\colhead{Earth A-band}
}

\startdata
2-m  & 1  & $1.35\times 10^5$ & $27000$            & $2490$ \\
     & 2  & $33700$           & $6760$             & $622$ \\
     & 5  & $5390$            & $1080$             & $100$ \\
10-m & 1  & $3.37\times 10^6$ & $6.76\times 10^5$  & $62200$ \\
     & 2  & $8.42\times 10^5$ & $1.69\times 10^5$  & $15500$ \\
     & 5  & $1.35\times 10^5$ & $27000$            & $2500$ \\
30-m & 1  & $3.03\times 10^7$ & $6.08\times 10^6$  & $5.6\times 10^5$ \\
     & 2  & $7.58\times 10^6$ & $1.52\times 10^6$  & $1.4\times 10^5$ \\
     & 5  & $1.21\times 10^6$ & $2.43\times 10^5$  & $22400$ \\
     & 50 & $12100$           & $2430$             & $224$ \\
 \enddata


\tablecomments{Table computed assuming flux ratios relative to the host star
$1.04\times 10^{-10}$ and $5.18\times 10^{-10}$ 
for Earth and Jupiter respectively, a Solar count rate of $2.6\times 10^{10}$~sec$^{-1}$
at 1~pc (appropriate to the F814W filter with ACS). The Earth A-band flux, 7580---7800\AA\ was
assumed to be 9.2\%\ of the F814W flux. An integration time of $10^4$~sec was assumed.}

\end{deluxetable}

\end{document}